\documentclass[traditabstract]{aa}  

\usepackage{natbib}

\usepackage{graphicx}
\usepackage{txfonts}
\def\degr{\hbox{$^\circ$}}
\def\arcmin{\hbox{$^\prime$}}
\def\arcsec{\hbox{$^{\prime\prime}$}}
\def\micron{\hbox{$\mu$m}}

\begin{document}

   \title{Global collapse of molecular clouds as a formation mechanism for the most massive stars}

   \author{N. Peretto
             \inst{1,2},
         G.~A. Fuller\inst{3,4},
         A.  Duarte-Cabral\inst{5,6}, A. Avison\inst{3,4}, P. Hennebelle\inst{1}, J.~E. Pineda\inst{3,4,7}, Ph. Andr\'e\inst{1}, S. Bontemps\inst{5,6}, F. Motte\inst{1}, N. Schneider\inst{5,6}, S. Molinari\inst{8}
                   }

   \institute{Laboratoire AIM, CEA/DSM-CNRS-Universt\'e Paris Diderot, IRFU/Service d'Astrophysique, C.E. Saclay, France 
                          \and    School of Physics and Astronomy, Cardiff University, Queens Buildings, The Parade, Cardiff CF24 3AA, UK\\
                            \email{Nicolas.Peretto@astro.cf.ac.uk}
                                    \and             Jodrell Bank Centre for Astrophysics, School of Physics and Astronomy, University of Manchester, Manchester, M13 9PL, UK
             \and                       UK ALMA Regional Centre node
             \and            Universit\'e de Bordeaux, LAB, UMR5804, F-33270, Floirac, France
             \and             CNRS, LAB, UMR5804, F-33270, Floirac, France
             \and             European Southern Observatory (ESO), Garching, Germany
             \and             IFSI, INAF, Area di Recerca di Tor Vergata, Via Fosso Cavaliere 100, I-00133, Roma, Italy
                                    }

   \date{Received 19 February 2013; accepted 2 June 2013}

 
  \abstract
   {
   The relative importance of primordial molecular cloud fragmentation versus large-scale accretion still remains to be assessed in the context of massive core/star formation. Studying the kinematics of the dense gas surrounding massive-star progenitors can tell us the extent to which large-scale flow of material impacts the growth in mass of star-forming cores. Here we present a comprehensive dataset of the 5500($\pm800$) M$_{\odot}$ infrared dark cloud SDC335.579-0.272 (hereafter SDC335) which exhibits a network of cold, dense, parsec-long filaments. Atacama Large Millimeter Array (ALMA) Cycle 0 observations reveal two massive star-forming cores, MM1 and MM2, sitting at the centre of SDC335 where the filaments intersect. With a gas mass of 545($^{+770}_{-385}$)~M$_{\odot}$ contained within a source diameter of $0.05$~pc, MM1 is one of the most massive, compact protostellar cores ever observed in the Galaxy. As a whole, SDC335 could potentially form an OB cluster similar to the Trapezium cluster in Orion. ALMA and Mopra single-dish observations of the SDC335 dense gas furthermore reveal that the kinematics of this hub-filament system are consistent with a global collapse of the cloud. These molecular-line data point towards an infall velocity  $V_{inf} =0.7(\pm0.2)$~km/s, and a total mass infall rate $\dot{M}_{inf}\simeq 2.5(\pm1.0)\times10^{-3}$~M$_{\odot}$\,yr$^{-1}$ towards the central pc-size region of SDC335. This infall rate brings $750(\pm300)$~M$_{\odot}$ of gas to the centre of the cloud per free-fall time ($t_{ff}=3\times10^5$~yr). This is enough to double the mass already present in the central pc-size region in  $3.5^{+2.2}_{-1.0} \times t_{ff}$. These values suggest that the global collapse of SDC335 over the past million year resulted in the formation of an early O-type star progenitor at the centre of the cloud's gravitational potential well.
   }
   
\titlerunning{Global collapse of the SDC335 massive star forming cloud}
\authorrunning{N. Peretto et al.}
   \maketitle
%

\section{Introduction}

The formation of massive stars remains, in many ways, a mystery \citep{beuther2007b,zinnecker2007}. More specifically, the key question of which physical processes determine their mass accretion history is yet to be answered. On one hand, some theories predict that primordial fragmentation of {\it globally stable} molecular clouds may form compact reservoirs of gas, called cores (with sizes up to 0.1pc), from which a forming star {\it subsequently} accumulates its mass \citep{mckee2003,beuther2004}. In an alternative scenario, molecular clouds {\it undergo global} collapse \citep{peretto2006,peretto2007}, gathering matter from large scales to the centre of their gravitational potential well, where cores, and protostars in them, are {\it simultaneously} growing in mass \citep{bonnell2004,smith2009}. Even though only statistical studies of large samples of massive star-forming clouds can definitely tell us which of these scenarios, if any, is most relevant, detailed observations of individual  massive star-forming clouds can still provide important hints.

   \begin{figure*}
   \centering
  \includegraphics[width=18cm]{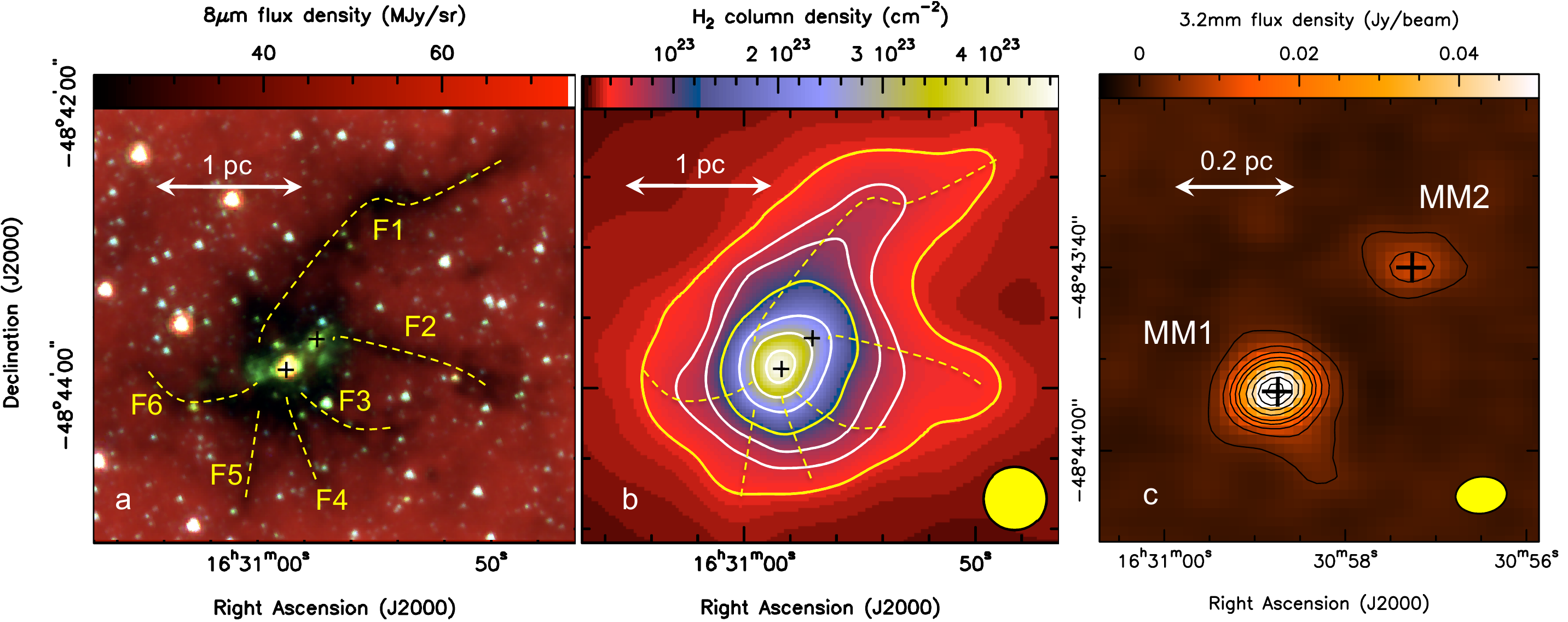}
    \caption{(a) Mid-infrared {\it Spitzer} composite image (red: 8\micron; green: 4.5\micron; blue: 3.6\micron) of SDC335. The 6 filaments identified by eye are indicated with yellow dashed lines, emphasizing their converging pattern. The diffuse 4.5\micron\ emission associated with the two IR sources in the centre is usually interpreted as a signature of powerful outflow activity. The positions of the two cores are marked with black crosses. (b) {\it Herschel} column density image of SDC335. The locations of the filaments and cores are marked similarly as in the (a) panel. The final angular resolution of this image is 25\arcsec\ (yellow circle), that of {\it Herschel} at 350\micron\ (see text). The contours range from $3.5\times10^{22}$ to $9.5\times10^{22}$~cm$^{-2}$ in steps of $2\times10^{22}$~cm$^{-2}$, and from $2.15\times10^{23}$ to $4.15\times10^{23}$~cm$^{-2}$ in steps of $1\times10^{23}$~cm$^{-2}$. The two yellow contours define the regions in which we calculated the SDC335 and Centre region masses quoted in Table 1. (c) ALMA 3.2mm dust continuum emission of the central region of SDC335 where two cores are identified, MM1 and MM2. The rms noise is 0.4~mJy/beam. The contours range from 2 to 22 in steps of 5 mJy/beam, and from 22 to 62 in steps of 10 mJy/beam. The yellow ellipse represents the ALMA beam size.}
              \label{spitalma}%
    \end{figure*}

The cloud under investigation is the {\it Spitzer} dark cloud SDC335.579-0.292 \citep[hereafter SDC335;][]{peretto2009}, a massive infrared dark cloud (IRDC) located at a distance of 3.25~kpc from the Sun (distance obtained using the \citealt{reid2009} model). The low level of radiative feedback from protostars in IRDCs ensures that the initial conditions for star formation are still imprinted in the gas properties \citep{rathborne2006,peretto2010a}. Massive IRDCs, such as SDC335, are therefore ideal places to study the earliest stages of high-mass star formation \citep[e.g.][]{kauffmann2010}. SDC335 exhibits a remarkable network of filaments seen in extinction at 8\micron\ (Fig.~\ref{spitalma}), reminiscent of hub-filament systems \citep{myers2009} observed in a number of low-mass \citep{andre2010,peretto2012} and high-mass \citep{schneider2012,hennemann2012} star-forming regions. The SDC335 filaments intersect at the centre of the cloud where two infrared protostars \citep[L$_{bol}>2\times10^4$~L$_{\odot}$;][]{garay2002} excite extended 4.5\micron\ emission, a tracer of powerful outflow activity \citep{cyganowski2008}. Consistently, class II methanol masers, unique tracers of massive star formation \citep{xu2008}, have also been reported towards these sources \citep{caswell2011}. However, despite these signposts of massive star formation, no 6cm free-free emission has been detected towards SDC335 down to a limit of 0.2 mJy \citep{garay2002}. This shows that little gas has been ionized in the centre of SDC335 and suggests that we are witnessing the early stages of the formation of, at least, two massive stars. 
 
The goal of this paper is to map the dense gas kinematics of SDC335 and analyse it in the context of massive star formation scenarios. In Section 2 we describe the observations. In Section 3 we discuss the mass partition in SDC335, and  Section 4 presents observations of the SDC335 dense gas kinematics. Finally, we discus our results and their implications in Section 5,  the summary and conclusions are presented in Section 6.


\section{Observations}

\subsection{{\it Spitzer} and {\it Herschel} observations}

We used publicly available\footnote{http://irsa.ipac.caltech.edu/data/SPITZER/GLIMPSE} {\it Spitzer} GLIMPSE data \citep{churchwell2009}.  The angular resolution of the 8\micron\ data is $\sim2$\arcsec. We also used the PACS \citep{poglitsch2010} 160\micron\ and SPIRE \citep{griffin2010} 350\micron\ {\it Herschel} \citep{pilbratt2010} data from the Hi-GAL survey \citep{molinari2010}. These data were reduced as described in \citet{traficante2011}, using the ROMAGAL map making algorithm. The nominal angular resolution at these two wavelengths are 12\arcsec\ and 25\arcsec.

\subsection{Mopra observations}

In May 2010 we observed SDC335 with the ATNF Mopra 22m single-dish telescope. We observed transitions including HCO$^+$(1-0), H$^{13}$CO$^+$(1-0) and N$_2$H$^+$(1-0) in a $5\arcmin\times5\arcmin$ field centred on SDC335. We performed on-the-fly observations, switching to an off-position free of dense gas emission. Pointing was checked every hour and was found to be better than 10\arcsec. We used the zoom mode of the MOPS spectrometer providing a velocity resolution of 0.1~km/s. The angular resolution of these 3mm Mopra observation is $\sim37\arcsec$ and the rms noise is 0.1~K  on the T$_A^*$ scale ($\sim 0.2$~K on the T$_{\rm{mb}}$ scale because the beam efficiency factor is $\sim2$ at 93~GHz on Mopra - \citealt{ladd2005}).

\subsection{ALMA  observations}

In September and November 2011 we observed SDC335 at 3mm wavelength with the 16 antennas of  ALMA (Cycle 0) in its compact configuration. We performed an 11-pointing mosaic covering the entire area seen in extinction with {\it Spitzer} (Fig.~\ref{spitalma}a). We simultaneously observed the 3.2mm dust continuum, along with the CH$_3$OH(13-12) and N$_2$H$^+$(1-0) transitions at a spectral resolution of $\sim0.1$~km/s. Flux and phase calibration were performed on Neptune and B1600-445, respectively.  The data were reduced using CASA\footnote{http://casa.nrao.edu} \citep{mcmullin2007}. The synthesized beam is $5.6\arcsec\times4.0\arcsec$ with a position angle of $+97\degr$. The rms noise in the continuum is 0.4~mJy/beam, while for the line we reach an rms sensitivity of 14~mJy/beam ($\sim0.08$~K).

As with any interferometer, ALMA filters out large-scale emission. To recover this emission, we used the Mopra single-dish data to provide the short-spacing information, for which we used the GILDAS\footnote{http://www.iram.fr/IRAMFR/GILDAS} software. This combination significantly improved the image quality,  in particular in the central region of SDC335. The rms noise on these combined datacubes is 0.14~Jy/beam ($\sim0.8$~K), significantly higher than the ALMA-only dataset. This reflects the higher noise of the Mopra dataset per ALMA beam.

\section{Mass partition in SDC335}

The mid-infrared composite image of SDC335 is displayed in Fig.~\ref{spitalma}a. In extinction we easily identify a network of six filaments (indicated as yellow dashed lines, and named F1 to F6), while in the centre of SDC335 we observe bright infrared sources exciting diffuse 4.5\micron\ emission (in green). In the following we provide mass measurements of the entire SDC335 clump, the filaments, and the cores at the centre.

\subsection{Clump and filaments}

\begin{table}
\caption{ Mass partition in SDC335. }
\centering
\begin{tabular}{c c c c c }
\hline\hline
Structure & Sizes & Mass            & Volume density  \\
  name        & (pc)    &   (M$_{\odot}$)& (cm$^{-3}$)        \\
\hline
SDC335 & 2.4 & $5.5( \pm0.8) \times10^3$ & $1.3(\pm0.2)\times10^4$   \\
Centre   &   1.2  & $  2.6( \pm0.3) \times10^3$  & $5.0(\pm0.6)\times10^4$  \\
F1	     &	$0.3\times2.0$	&  $0.4 (\pm0.1) \times10^3$& $4.9(\pm1.3)\times10^4$  \\
F2          &  $0.3\times1.3$  & $0.2 (\pm0.1) \times10^3$&  $3.7(\pm1.8)\times10^4$  \\
MM1      &   0.054  & $545(^{+770}_{-385})$  & $1.1(^{+1.7}_{-0.8})\times10^8$  \\
MM2      &   0.057  & $65(^{+92}_{-46})$    & $1.2(^{+1.6}_{-0.9})\times10^7$    \\
\end{tabular}
\tablefoot{The sizes are all beam deconvolved and  refer either to diameters when spherical geometry is assumed, or to diameters  $\times$ lengths  when cylindrical geometry is assumed.}
\label{tabmass}
\end{table}

Mid-infrared extinction mapping of IRDCs is a powerful method for measuring their column density distribution at high resolution \citep{peretto2009,butler2009}. From the resulting maps one can measure the masses of these clouds. However, this method is limited by definition to mid-infrared absorbing dust, and in cases such as SDC335, where bright 8\micron\ sources have already formed, the resulting dust extinction masses become more uncertain. Using {\it Herschel} data allows us to circumvent this problem by looking at far-infrared dust emission from 70\micron\ up to 500\micron. A standard way for recovering the column density distribution from {\it Herschel} data is to perform a pixel-by-pixel spectral energy distribution fitting after smoothing the data to the {\it Herschel }500\micron\ resolution (36\arcsec). To obtain a higher angular resolution, we decided here to use the 160\micron/350\micron\ ratio map\footnote{Note that we did not use the 250\micron\ image because of the significant fraction of saturated pixels at the centre of   SDC335.} of SDC335 as an indicator of the dust temperature, and then reconstruct the column density distribution at 25\arcsec\ resolution by combining the dust temperature ($T_d$) and 350\micron\ ($S_{350}$) maps, assuming that dust radiates as a modified black-body. The column density is therefore written as:
\begin{equation}
N_{H_2}(x,y)=S_{350}(x,y)/[B_{350}(T_d(x,y))\kappa_{350}\mu m_H]
\end{equation}
where $B_{350}$ is the Planck function measured at 350\micron, $\mu=2.33$ is the average molecular weight, $m_H$ is the atomic mass of hydrogen, and $\kappa_{350}$ is the specific dust opacity at 350\micron. 
Using the dust opacity law \citep{hildebrand1983,beckwith1990}  $\kappa_{\lambda}=0.1\times\left(\frac{\lambda}{0.3mm}\right)^{-\beta}$~cm$^2$g$^{-1}$ with $\beta=2$, we constructed the SDC335 column density map presented in Fig.~\ref{spitalma}b. We can see that, despite the difference in resolution, the column density structure of SDC335 follows the extinction features we see in the mid-infrared. Moreover, the entire central pc-size region lies above a high column density of $1\times10^{23}$~cm$^{-2}$. 

The dust opacity parameters we used to construct the SDC335 column density map are very uncertain. There are indications \citep[e.g.][]{paradis2010} that for cold, dense clouds a spectral index $\beta\simeq2.4$ is possibly more appropriate. The net effect of using a lower $\beta$ is to overestimate the temperature, and therefore underestimate the column density (and the mass calculated from this column density). Also, the dust emission towards IRDCs is composed of the emission from the the cloud itself and emission from the galactic plane background/foreground dust, which is warmer and more diffuse. A consequence of this is that, once again, the IRDC temperature we calculate is overestimated. The measured average dust temperature towards the coldest parts of SDC335 is $\sim 16$~K, which  appears to be, indeed, 2-3~K  warmer compared with other IRDCs \citep{peretto2010b}. Altogether, the column densities presented in Fig.~\ref{spitalma}b are likely to be underestimated. To obtain an upper limit on the SDC335 column density we decreased the dust temperature map by 2~K, which relocates SDC335 in the typical temperature range observed in IRDCs. We then recalculated the column density map using Eq.~(1). Masses and uncertainties quoted in Table~\ref{tabmass} for SDC335, the Centre region, and the F1 and F2 filaments were obtained taking the average mass obtained from the two {\it Herschel} column density maps. All masses are background subtracted, which corresponds to the mass enclosed in a specific column density contour for a centrally concentrated source.  For SDC335 and the Centre region, the sizes correspond to the diameter of the disc with the same areas, for F1 and F2 they correspond to the two dimensions of the rectangle with the same areas. The surface areas of these two filaments correspond to the polygons we drew on the {\it Herschel} column density map around the F1 and F2 filaments. The widths of these polygons are constrained by the filament profiles as seen in the 8\micron\ extinction  map (4\arcsec\ resolution), which correspond to $\sim0.3$~pc. Note that for the filaments we performed an independent mass measurement directly using the 8\micron\ extinction map from \citet{peretto2009}, confirming the {\it Herschel} mass measurements. Densities were then calculated assuming a uniform density and spherical geometry for SDC335 and the Centre region, and cylindrical geometry for the F1 and F2 filaments.

\subsection{Dense cores}

The ALMA 3.2mm dust continuum observations presented in Fig.~\ref{spitalma}c show two bright sources, MM1 and MM2, which have J2000 coordinates (RA: 16$^h$30$^m$58.76$^s$; Dec: -48\degr43\arcmin53.4\arcsec) and (RA: 16$^h$30$^m$57.26$^s$; Dec: -48\degr43\arcmin39.7\arcsec), respectively. Each of these cores is associated with one infrared source and class II methanol maser spots \citep{caswell2011}, leaving no doubt that MM1 and MM2 are currently forming massive stars.

\begin{figure}
   \vspace{-0cm}
   \hspace{0.5cm}
   \includegraphics[width=8cm,angle=0]{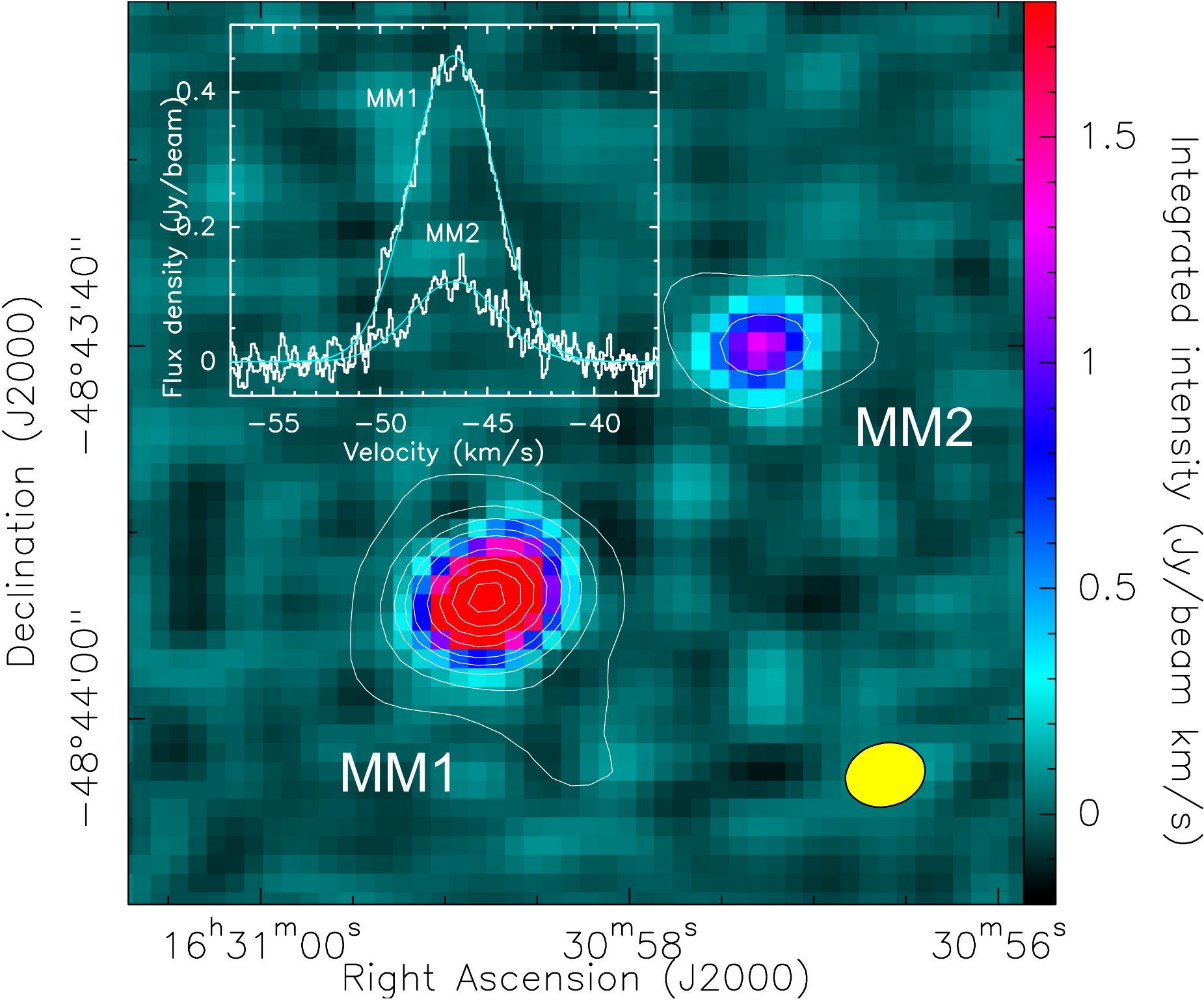}
   \vspace{-0cm}
     \caption{ALMA CH$_3$OH(13-12) integrated-intensity image of SDC335 in colour scale; overplotted are the contours of the 3.2mm dust continuum emission as displayed in Fig.~\ref{spitalma}c. We can see that the two types of emission coincide spatially . The insert in the top left corner shows the methanol spectra observed at the central position of each core. The cyan solid lines are the best Gaussian fit to the data. The yellow ellipse represents the ALMA beam size.
      }
         \label{ch3oh}
   \end{figure}

The MM1 and MM2 cores are compact but partially resolved. The results of a 2D Gaussian fit to the 3.2mm continuum emission of these compact sources give integrated fluxes of 101($\pm10$)~mJy and 12($\pm2$)~mJy and deconvolved sizes of 0.054~pc and 0.057~pc for MM1 and  MM2, respectively. To estimate how much of this emission could be free-free, we scaled the 3-$\sigma$ non-detection limit at 6cm from \citet{garay2002} to 3.2mm using $F^{\lambda}_{\rm   ff}= 0.2~\rm{mJy}\times [60/\lambda(mm)]^{\alpha}$. Upper limits on $\alpha$ have been determined for a set of massive protostellar objects and HCHII regions \citep{cyganowski2011,hoare2005} consistent with $\alpha=1$\footnote{Note that in the same   study the authors also determine a lower limit of  $\alpha>1.7$ for one source, but this measurement is based on a single 4.2$\sigma$   detection of a very weak source.}. This provides an upper limit for free-free contamination at 3.2mm of $F^{3.2}_{\rm ff}=4$ mJy. Clearly, this is negligible for MM1, whereas it could contribute up to 33\% of the MM2 flux. In the strict upper-case limit of optically thick free-free emission at both 6cm and 3.2mm  $\alpha$ equals 2, increasing the upper limit for free-free contamination to 70\% of the MM1 ALMA 3.2mm flux. However, the recent detection of MM1 at 7mm with ATCA (Avison et al. in prep.) shows no excess emission over that expected from the dust. Still, conservatively assuming that the entire 7mm emission is from optically thick free-free emission implies an upper limit on the free-free contamination of 30\% of the observed 3.2mm flux. Such a level of contamination would not change any of the results presented here, therefore we neglect any potential free-free contamination in the remainder of the paper. The gas mass and the 3.2mm flux of the MM cores are related through
\begin{equation} 
M_{gas}=\frac{d^2F_{3.2}}{\kappa_{3.2}B_{3.2}(T_d)} \, ,
\end{equation}
where $d$ is the distance to the source, $F_{3.2}$ is the 3.2mm flux, $\kappa_{3.2}$ is the specific dust opacity (accounting for the dust-to-gas-mass ratio) and $B_{3.2}(T_d)$ is the Planck function measured at 3.2mm with a dust temperature $T_d$. The main sources of uncertainties on this mass estimate come from the dust properties, temperature, and opacity.  The dust temperatures of these two sources are difficult to determine based on these ALMA observations alone. However, both sources have strong mid/far-IR emission seen with {\it Spitzer} \citep{churchwell2009} and {\it Herschel} \citep{molinari2010}, class II methanol maser emission \citep{caswell2011}, and are detected in high-excitation thermal lines of methanol (Sec. 4).  These are indicative of dust in the centre of the cores with temperatures $>100$~K, but it is also clear that on larger scales the dust within the dark SDC335 filaments is cold, with temperatures $\sim15$~K as measured in many other IRDCs \citep{peretto2010b,wilcock2012}. For the vast majority of massive protostellar cores in the literature (cf caption of Fig.~\ref{radmass}), the assumed or measured dust/gas temperature (via SED or K-ladder fitting of some specific lines) varies between 15~K and 100~K. Here, we adopted an intermediate dust temperature of 50~K for both MM cores, and considered that a factor of 2 uncertainty on this dust temperature is conservative. In the future, radiative transfer modelling of these sources will be necessary to better constrain their temperature profiles.

We took the same dust opacity law as used for the {\it Herschel} data, providing $\kappa_{3.2}=8.7\times10^{-4}$~cm$^2$g$^{-1}$ (assuming a dust-to-gas-mass ratio of 1\%). However, this value is sensitive to the dust model used. It is unclear which model is the most appropriate for protostellar cores, but as shown in Fig.~\ref{radmass}, most values adopted in the literature for core-mass measurements agree within a factor of 2. With these assumptions we estimated the gas masses and their associated uncertainties as $M_{MM1}=545^{+770}_{-385}$~M$_{\odot}$ and $M_{MM2}=65^{+92}_{-46}$~M$_{\odot}$.

 \begin{figure}
   \vspace{-0cm}
   \hspace{0.5cm}
   \includegraphics[width=7.3cm,angle=0]{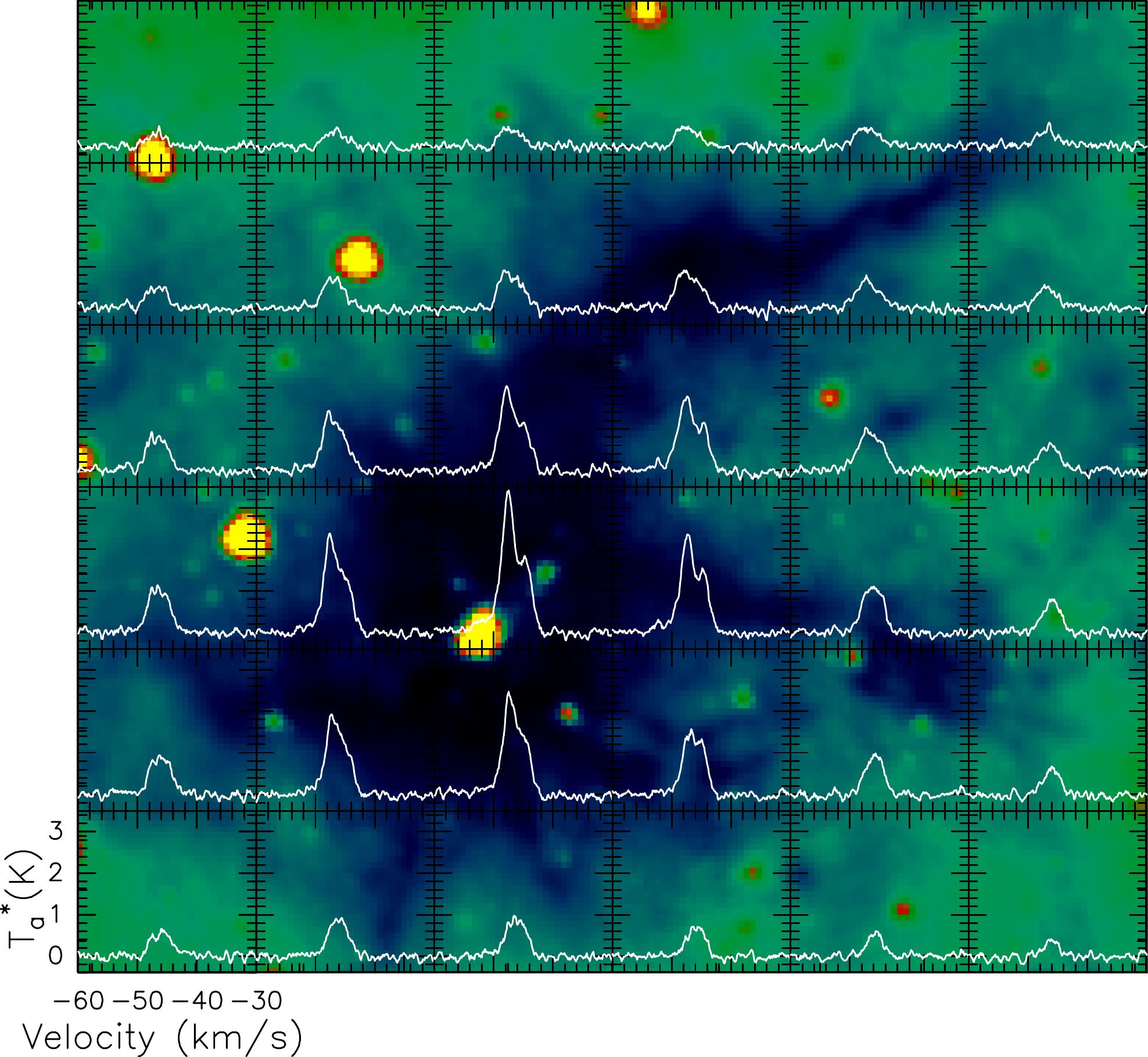} 
   \vspace{-0cm}
    \caption{Spitzer 8\micron\ image of SDC335 (colour scale) over-plotted with the Mopra HCO$^+$(1-0) spectra. The temperature scale and velocity are indicated in the bottom-left corner. The HCO$^+$(1-0) line is self-absorbed and blue-shifted in the bulk of the cloud. This is usually interpreted as a signature of collapse.}
         \label{hcop}
   \end{figure}

 \begin{figure*}
\centering
\includegraphics[width=18.5cm]{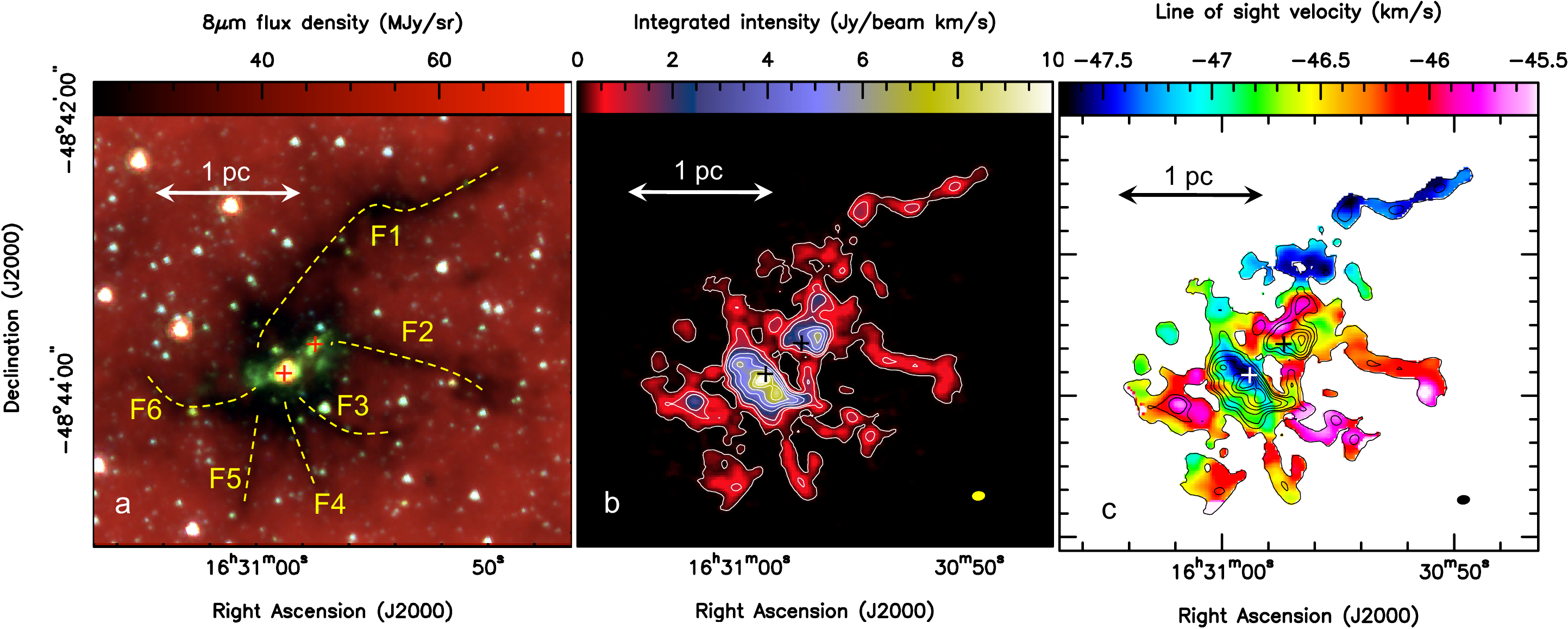}
 \caption{(a) Same as in Fig.~\ref{spitalma}a; (b) ALMA-only image of the N$_2$H$^+$(1-0) integrated intensity over the 7 hyperfine structure components. The rms noise on the resulting map is $\sim 6$~mJy/beam\,km/s. The contours go from 0.1 to 1.5 in steps of 0.7 Jy/beam\,km/s and 1.5 to 9 in steps of 1.5 Jy/beam\,km/s. The crosses mark the positions of the two dense cores. The ALMA beam is represented as a yellow elliptical symbol in the bottom-right corner of the image. We can see the excellent match between the Spitzer dust extinction of the filaments and the N$_2$H$^+$(1-0) emission; (c) ALMA N$_2$H$^+$(1-0) velocity field using the first order moment map. The crosses mark the positions of the cores and the contours are the same as in the (b) panel. }
              \label{n2hp}%
    \end{figure*}

\section{Dense-gas kinematics in SDC335}

In this section we discuss the dense-gas kinematics of the cores, filaments, and clump as observed with the Mopra and ALMA telescopes.

\subsection{ALMA CH$_3$OH(13-12) core velocities}

To determine the systemic velocity of the MM cores, we mapped the thermal methanol CH$_3$OH(13-12) transition at 105.063761 GHz. Because of the high energy levels of this transition (E$_{u}=223.8$~K), CH$_3$OH(13-12) is preferentially observed in dense and warm regions. Figure~\ref{ch3oh} shows the ALMA integrated-intensity image towards the cores. We see the excellent agreement between the position of the peak of the dust continuum cores and the methanol emission, indicating that methanol is a good tracer of their systemic velocities. We also note that the methanol emission is more compact (unresolved, i.e  $<0.01$~pc) than the dust continuum emission, which indicates that it arises from the warm, innermost regions of the cores.  Gaussian fits to the methanol spectra observed at the central position of the two cores (insert of Fig.~\ref{ch3oh}) provide the systemic velocities of the cores (V$_{MM1}=-46.6$km/s , V$_{MM2}=-46.5$km/s) and the gas velocity dispersion in the densest parts of MM1 and MM2 ($\Delta V_{MM1}=4.6$km/s , $\Delta V_{MM2}=4.8$km/s).

\subsection{Mopra HCO$^+$(1-0) self-absorbed lines}

HCO$^+$ is a well-known tracer of dense gas in molecular clouds.
 In these regions,  HCO$^+$(1-0) can be optically thick,  in which case the line shape can provide information of the global motions of the gas along the line of sight \citep[e.g. ][]{Fuller2005,smith2012}. The HCO$^+$(1-0) observations towards SDC335 (Fig.~\ref{hcop}) show blue-shifted self-absorbed spectra in the bulk of the cloud. Such line profiles are expected for an optically thick tracer of idealized collapsing clouds in which the excitation temperature is rising towards the centre. What is important to note here is the extent (over at least 12 independent Mopra beams) over which this spectral signature is observed, and  the absence of any other line asymmetry. For expanding motions we would expect red-shifted self-absorbed spectra, while in the case of rotation blue-shifted and red-shifted spectra on either side of the rotation axis should be produced. Therefore these  HCO$^+$(1-0) observations towards SDC335 already rule out the possibility of a rotating or expanding cloud, and strongly suggest that SDC335 is collapsing. 
 
 SDC335 is  well enough characterised that we can estimate the HCO$^+$ abundance using the 1D non-LTE RADEX radiative transfer code \citep{vandertak2007}. This code predicts line intensities based on a set of input parameters for which we have strong constraints: the kinetic temperature ($20\pm5$~K, estimated from {\it Herschel} data), the cosmic background temperature (2.73~K), the central H$_2$ density averaged over the Mopra beam ($6\pm1\times10^4$~cm$^{-3}$, estimated from the column density map presented in Fig.~\ref{spitalma}), and the velocity dispersion ($1.3\pm0.3$~km/s; cf Sect.~5.4). Then we iterate on the last input parameter, i.e. the molecule column density, to match the model line intensities with the observed line temperature, i.e. $T_{HCO^+}^{peak}=6.4(\pm0.2)$~K on the T$_{mb}$ scale.  Doing so, we obtain $N_{HCO^+}=6^{+7}_{-3}\times10^{13}$~cm$^{-2}$, corresponding to an abundance $X_{HCO^+}=7^{+8}_{-4}\times10^{-10}$. The corresponding excitation temperature is $T_{ex}=10.4^{+1.2}_{-0.7}$~K, confirming that HCO$^+$(1-0) is not thermalised. Using the same set of parameters, we performed the same exercise for the central H$^{13}$CO$^+$(1-0) line (se Fig.~\ref{h13copmodel}), which has  $T_{H^{13}CO^+}^{peak}=1.2(\pm0.2)$~K on the T$_{mb}$ scale. For this line we obtain $N_{H^{13}CO^+}=4^{+3}_{-2}\times10^{12}$~cm$^{-2}$, corresponding to an abundance $X_{H^{13}CO^+}=5^{+3}_{-3}\times10^{-11}$. The corresponding excitation temperature is $T_{ex}=6.5^{+2.7}_{-1.2}$~K. Therefore, as for HCO$^+$(1-0), H$^{13}$CO$^+$(1-0) is not thermalised. Note that the lower excitation temperature of H$^{13}$CO$^+$(1-0) is most likely due to a lower beam-filling factor. Another important point is  that given the abundances we calculated for both molecules, we obtain an abundance ratio $15\leq[\rm{HCO^+]/[H^{13}CO^+}] \leq20$. The $\rm{[^{12}C]/[^{13}C]}$ ratio is known to increase as a function of the galactocentric radius, and at  the galactocentric distance of SDC335 (i.e. $\sim5$~kpc)  the predicted  $\rm{[^{12}C]/[^{13}C]}$ is around 30 \citep{langer1993,savage2002}. The value we find is about half this value, which, considering the uncertainties on these kinds of measurements, is in reasonable agreement. We use the latter value of the fractional abundance for the radiative modelling presented in Sect.~5.3.

\begin{figure}
\centering
\includegraphics[width=7cm]{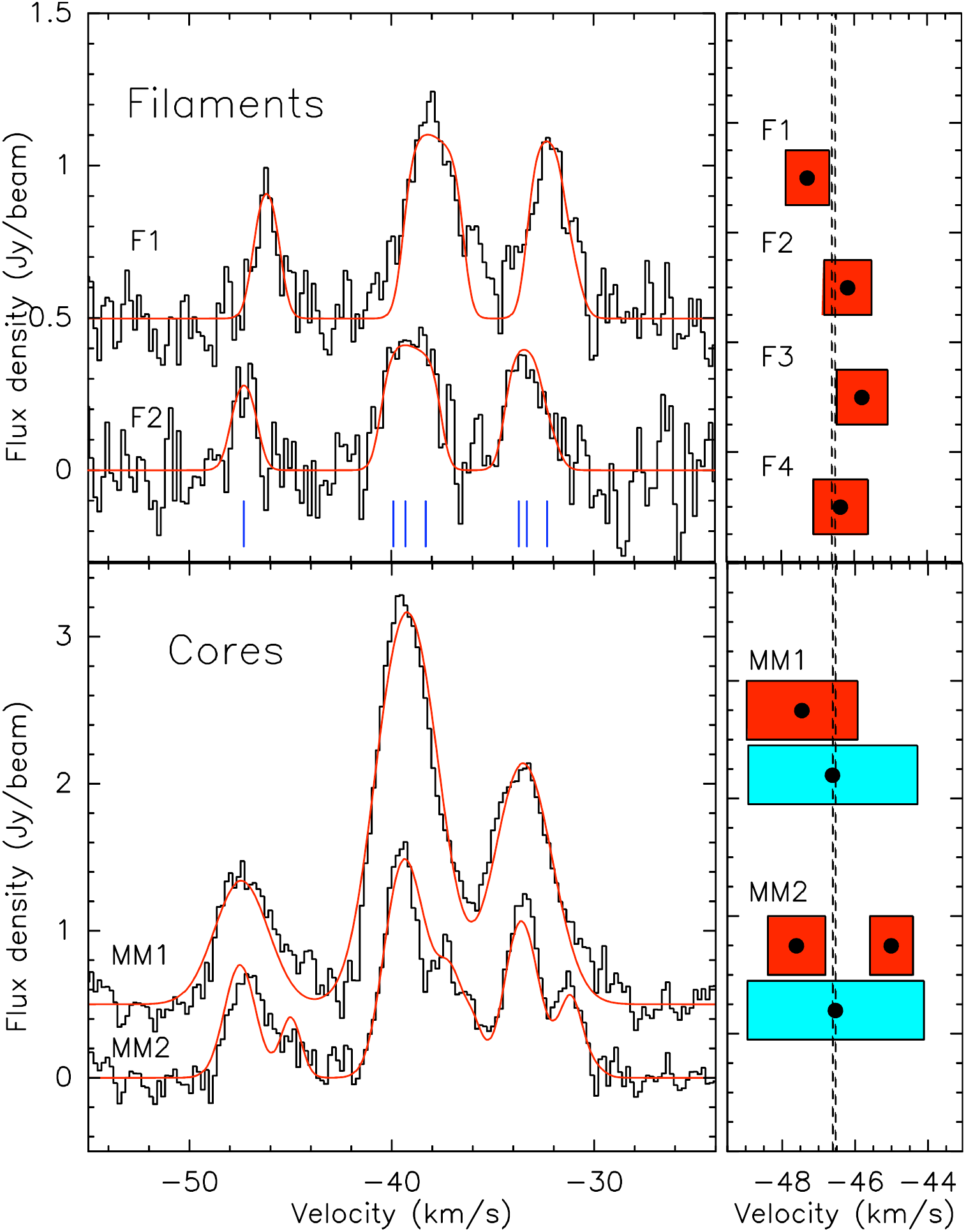}
 \caption{ (Left) Examples of combined ALMA and Mopra N$_2$H$^+$(1-0) spectra observed at some specific positions in SDC335 (upper panel for filaments - lower panel for cores), along with their best fits as red solid lines. The N$_2$H$^+$(1-0) spectra exhibit a hyperfine structure (HFS) composed of 7 components (the positions are displayed as blue vertical ticks for the F2 filament). Some of these components are close enough to be blended when the velocity dispersion of the gas is supersonic, resulting in three groups of lines. For a kinetic temperature of 10K, the velocity dispersion along the filaments is supersonic by a factor 1.5 to 3, similar to what is observed in other IRDCs \citep{ragan2012}.
 (Right) Schematic representation of the systemic velocity and velocity dispersion of the different structures. The length of each box represents the velocity dispersion (FWHM) of the gas, and its central position the systemic velocity (represented as a filled circle). The colour of the boxes codes the line which has been used for the measurements: red for N$_2$H$^+$(1-0), and cyan for CH$_3$OH(13-12). The vertical dashed lines mark the systemic velocities of the cores.}
              \label{spectra}%
    \end{figure}

\subsection{ALMA N$_2$H$^+$(1-0) cloud velocity field}

Figure~\ref{n2hp}b shows the ALMA N$_2$H$^+$(1-0) integrated-intensity map of SDC335. The visual comparison with the {\it Spitzer} image of SDC335 demonstrates how efficient this molecule is in tracing the network of pc-long filaments seen in dust extinction. This justifies our choice of using this line to trace the filaments kinematics. On the other hand, we can also see that N$_2$H$^+$ is a poor tracer of the cores, where the central heating may have partly removed it from the gas phase \citep{zinchenko2009,busquet2011}.

Figures~ \ref{n2hp}c and \ref{spectra} show that SDC335 velocity field is homogeneous in each filament, with distinct velocities from filament to filament (e.g. $<V_{F1}>=-47.4 \pm 0.1$ km/s; $<V_{F3}>=-45.8 \pm 0.2$ km/s). It becomes more complex towards the centre of the cloud. In Fig.~\ref{spectra} we see that two separate velocity components are present close to MM2, while the broad asymmetric line profiles around MM1 suggest their blending, consistent with observations of other massive cores \citep{csengeri2011}. This line shape cannot be the result of high optical depth since the N$_2$H$^+$(1-0) hyperfine line-fitting (performed with GILDAS) gives an opacity lower than 1 everywhere in the cloud. Kinematically, the gas traced by N$_2$H$^+$(1-0) at the centre of the cloud appears to be composed of a mix of the gas originating from the two main filaments, F1 and F2. Figure~\ref{spectra} (right) presents a schematic view of the velocities of the filaments and cores. We can actually see that  the two cores lie at an intermediate velocity between the velocities of the different filaments. This configuration suggests that the cores are at least partly fed by the pristine gas flowing along these filaments at a velocity $V_{inf} \simeq$ 1~km/s.

\section{Discussion}

\subsection{SDC335: An OB cluster progenitor}

\begin{figure}
   \vspace{-0cm}
   \hspace{0.5cm}
    \includegraphics[width=8cm,angle=0]{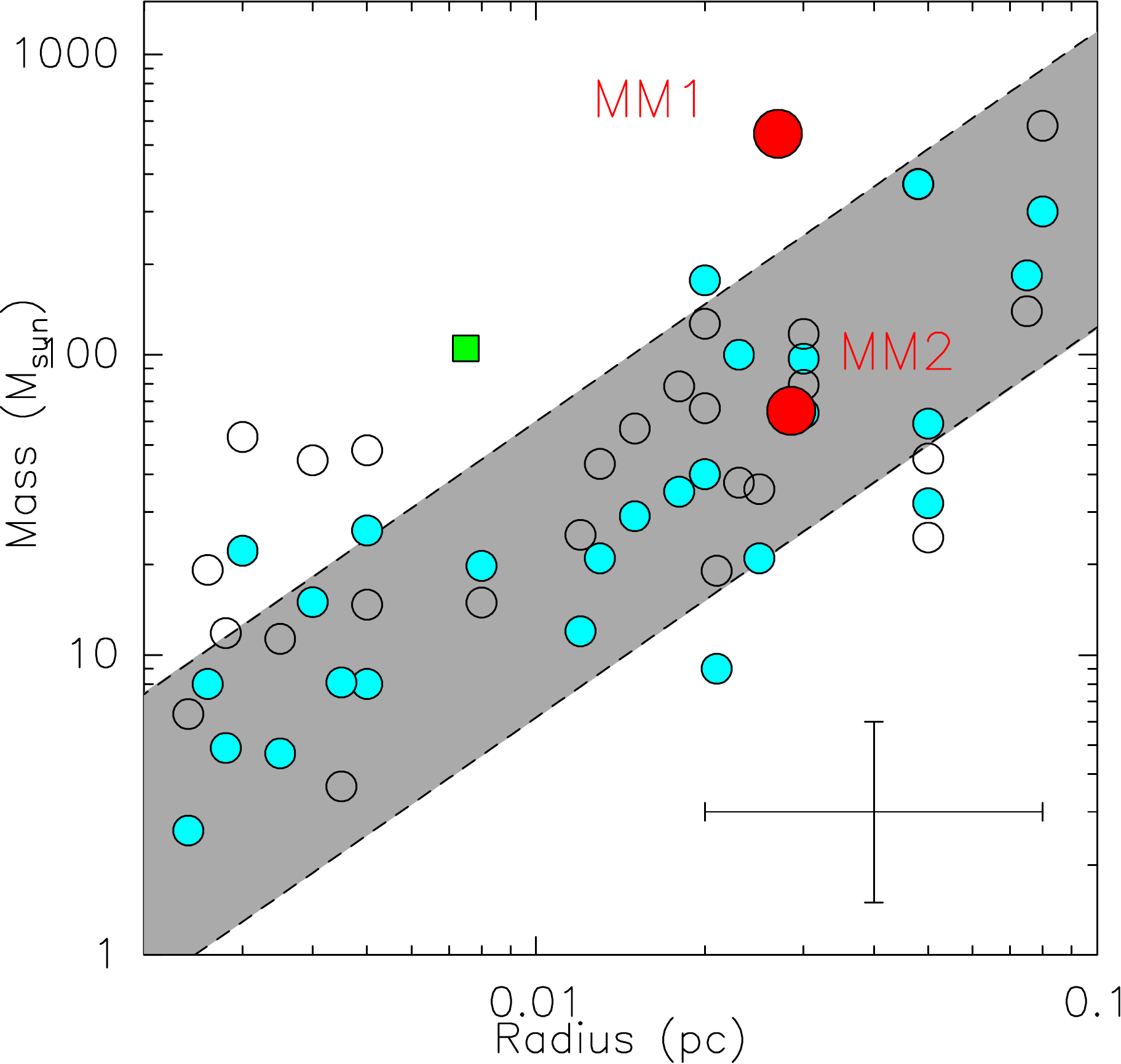}
   \vspace{-0cm}
     \caption{ Mass-radius diagram of massive protostellar cores. The cyan circles correspond to the values as published in the literature \citep{peretto2007,ren2012,rathborne2011,wang2011,zhang2011,bontemps2010,rathborne2007,rathborne2008,beuther2002,beuther2003,beuther2005,beuther2012,molinari1998,rodon2012}, while the empty circles correspond to the same set of sources for which we recalculated the mass using the same dust opacity law as in this paper. The MM1 and MM2 sources are indicated as red filled circles. The green filled square marks the position of the W51 North star+disc system from \citet{zapata2009}. The shaded area indicates the region where most sources lie. The cross in the bottom right corner indicates a factor of 2 uncertainty in both the masses and sizes, typical of  the results presented here.    
           }
         \label{radmass}
   \end{figure}

In Section~3 we inferred core masses of $M_{MM1}=545^{+770}_{-385}$~M$_{\odot}$ and $M_{MM2}=65^{+92}_{-46}$~M$_{\odot}$ in deconvolved diameters $0.05-0.06$~pc. Figure~\ref{radmass} shows a radius-versus-mass diagram for a significant (but not complete) sample of massive protostellar cores published in the literature. In this diagram we can see that SDC335~MM1 stands out, and for cores with similar sizes MM1 is a factor of between 3 and 20 more massive. However, given the uncertainties on the dust properties and density profile of cores, MM1 could match the mass of the most massive cores observed, on smaller scales, in Cygnus X \citep{bontemps2010}. Nevertheless, MM1 clearly appears to be one of the most massive, compact protostellar cores ever observed in the Galaxy.

Another interesting source that provides an informative comparison is W51 North. This source is believed to contain an already formed $\geq 65$~M$_{\odot}$ star, with a surrounding 3000~AU accreting disc of 40~M$_{\odot}$ \citep{zapata2009}. Adding this source to Fig.~\ref{radmass} using its total (star+disc) mass shows it to be an extreme object as well. SDC335~MM1 could represent an earlier version of such an O-type star-forming system.
For compact cores, the fraction of mass likely to be accreted onto the star is typically 50\% of the total core mass \citep[][]{duartecabral2013,mckee2003}. Despite probable unresolved fragmentation on smaller scales, the MM1 core and its large mass have the potential to form at least one star of 50~M$_{\odot}$ to 100~M$_{\odot}$.  

Assuming now that within SDC335 (M$=5500 \pm800$ M$_{\odot}$) a fully sampled standard initial mass function forms \citep{kroupa2002,chabrier2003},  then, in addition to the early O-type star in MM1, a thousand solar mass cluster of $\sim320$ stars  with masses from 1 to 50~M$_{\odot}$  should emerge from SDC335. Including lower mass stars in this calculation, we would reach a star formation efficiency $\geq 50\%$, the necessary condition to form a bound open cluster \citep[e.g.][]{lada2003}. As a whole, SDC335 could potentially form an OB cluster similar to the Trapezium cluster in Orion \citep{zinnecker2007}.

\subsection{A large mass reservoir for MM1}

We can estimate the conditions  under which MM1 formed within the context of gravo-turbulent fragmentation models. Using the standard (lognormal) volume-density probability density function (PDF) of non-self-gravitating turbulent clouds \citep[e.g.][]{padoan1997,hennebelle2008}, we calculated (see Appendix A) that less than 0.01\% of the gas is expected to be above a density of 10$^7$~cm$^{-3}$, while $>10\%$ of the SDC335 mass is above this threshold in the form of cores (see Table~1). Therefore, gravity must have brought together such a large mass in such a small volume. A first possibility is that the material  currently lying in MM1 was initially part of a larger volume that then collapsed. To calculate the diameter D$_{ini}$ of this volume we first need to calculate the density $\rho_{ini}$ above which 10\% of the gas is lying, assuming a lognormal density PDF. Using the observed parameters of SDC335 (see Appendix A), we found that D$_{ini}$ of this initial volume must have been $\sim15$ times larger than the current MM1 size, which means D$_{ini}\sim0.8$~pc. This size is in fact a lower limit since the calculation implicitly assumes that the entire dense gas above $\rho_{ini}$ lies within a single dense region. The second possibility is that MM1 initially had the same diameter as observed today. It is then possible to calculate the maximum mass that this volume can contain to match the lognormal PDF. We can  show (see Appendix A) that the maximum initial mass of such a core is  $\sim3$~M$_{\odot}$. This low mass means that most of the current MM1 mass must have been accreted from its surroundings. Using the current average density of SDC335, we calculated that the region from which MM1 accreted matter would had to have a diameter of 1.2pc. Either of these scenarios, therefore, requires large-scale, rather than local, accretion/collapse to form MM1. 

\subsection{Collapse on large scales}

\begin{figure}[!t!]
   \vspace{-0.cm}
   \hspace{0cm}
   \includegraphics[width=9cm,angle=0]{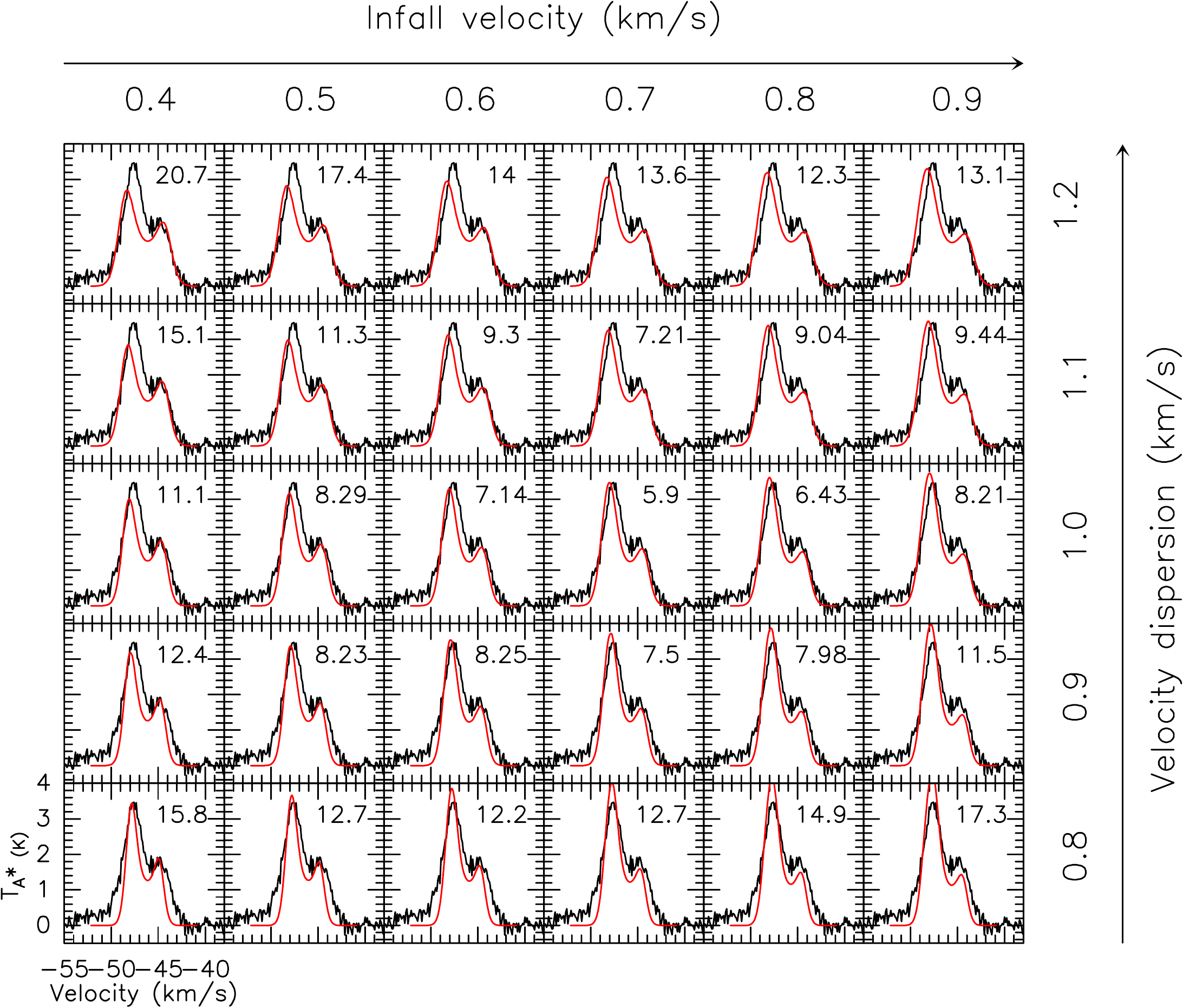} 
   \vspace{-0.5cm}
    \caption{Grid of HCO$^+$(1-0) spectra obtained from RATRAN modelling of a collapsing cloud (see text). All input parameters are fixed with the exception of the infall velocity ($V_{inf}$) and velocity dispersion ($\sigma_{turb}$). Each modelled spectrum (in red) has been obtained for the corresponding $V_{inf}-\sigma_{turb}$ displayed in the top and right-hand sides of the figure. The numbers in the top right-hand corner of each spectrum give a relative idea of the fit quality. The  HCO$^+$(1-0) spectrum observed at the centre of SDC335 is plotted in black. Since we aimed to keep the observed spectra displayed in T$_a^*$ scale, we applied a 0.5 factor to the modelled spectra to take into account the Mopra main-beam efficiency.}
         \label{specmodhcop}
   \end{figure}

The Mopra HCO$^+$(1-0) spectra presented in Fig.~\ref{hcop} are suggestive of global gravitational collapse. A simple analytical model \citep{myers1996} allows one to infer an infall velocity from these spectra. Using this model, we obtained an infall velocity of $\sim0.4$ km/s. However, as noted by \citet{devries2005}, this model underestimates the infall velocity by a factor of $\sim2$. We therefore decided to run a more sophisticated radiative transfer model to better constrain this infall velocity. For this purpose we used the RATRAN 1D Monte Carlo radiative transfer code \citep{hogerheijde2000}. The input parameters for the calculations are the mass of the cloud, its radius, density profile, kinetic temperature profile, turbulent velocity dispersion, the infall velocity profile, and the abundance profile of the line to be modelled. Obviously, a 1D model cannot describe the detailed kinematics of the filamentary structures observed in SDC335, and for this reason we decided to model only the central HCO$^+$(1-0) and H$^{13}$CO$^+(1-0)$ spectra. We used the SDC335 mass and size quoted in Table 1, a cloud-density profile  $\rho\propto r^{-1.5}$, and a constant temperature profile of 20~K. Based on the discussion in Sec.~4.2, we fixed the HCO$^+$ abundance (relative to H$_2$) to $7\times10^{-10}$  and an abundance ratio [HCO$^+$]/[H$^{13}$CO$^+$] of 30. We then ran a grid of models varying the last two input parameters, i.e., the infall velocity  and the velocity dispersion. Figure~\ref{specmodhcop} shows the results of the HCO$^+$(1-0) modelling of the central pixel for infall velocities ranging from 0.4~km/s to 0.9~km/s, and velocity dispersions from 0.8~km/s to 1.2~km/s. For each spectrum we calculated a reduced $\chi^2$ parameter representative of the quality of the fit. This parameter is given in the top right-hand corner of each spectrum. The corresponding H$^{13}$CO$^+$(1-0) predictions from the model are displayed in Appendix B. From Fig.~\ref{specmodhcop} we consider that $0.5~\rm{km/s} \leq V_{inf}\leq 0.9~\rm{kms}$ and  $0.9~\rm{km/s} \leq \sigma_{turb} \leq 1.1~\rm{km/s}$ provide reasonable fits to the central HCO$^+$(1-0) spectrum. We also performed models varying the radius of the collapsing region $R_{inf}$. For $R_{inf} < 0.5$~pc the modelled HCO$^+$(1-0) spectra remain symmetric, which is inconsistent with the observations. Only for $R_{inf}\geq 0.8 $pc is the asymmetry large enough to resemble the observed one. This shows that the observed HCO$^+$(1-0) self-absorbed spectra towards SDC335 do trace global collapse.

\subsection{Energy balance}

To be collapsing, the gravitational energy of a cloud has to overcome the kinetic energy that supports it. This occurs if the virial parameter $\alpha_{vir}=5\sigma^2_{turb}R/(GM)$ is lower than 1 \citep{bertoldi1992}. In this equation $\sigma_{turb}$, R, and M are the 1D velocity dispersion, the cloud radius, and the cloud gas mass. In the case of SDC335, we estimated $\sigma_{turb}=1.3$~km/s from the averaged N$_2$H$^+$(1-0) spectrum over SDC335 as observed with Mopra. Using Mopra $^{13}$CO(1-0) data, which trace less dense gas, we obtained $\sigma_{turb}=1.6$~km/s. Note also that these velocity-dispersion measurements include any systematic motions within the beam, such as infall, which artificially increase the velocity-dispersion estimate \citep{peretto2007}. Taking this into account, and because the filaments are well traced by N$_2$H$^+$, we estimate $\sigma_{turb}=1.3(\pm0.3)$~km/s. With R=1.2~pc and $M=5500(\pm800)$~M$_{\odot}$ we find $\alpha_{vir}=0.4_{-0.2}^{+0.4} < 1$. Additional support against gravity could be provided by the magnetic field. Following previous studies \citep{pillai2011}, we estimated that the magnetic field strength $|B|_{vir}$ necessary to virialize SDC335 is $|B|_{vir}=300~\mu$G, which is at least three times higher than observations of clouds at similar densities suggest \citep{crutcher2012}. Finally, note that the support provided by centrifugal forces can potentially stabilise a cloud against gravity. However, calculating the rotational energy of SDC335 by assuming that it is a homogenous rotating sphere with an angular velocity $\omega=1$~km/s/pc, we estimated that it is $\sim10$ times smaller than its kinetic energy as measured from the velocity dispersion. In other words, it is negligible.

\subsection{Large-scale velocity field and accretion rates}

\begin{figure}
\centering
\includegraphics[width=9cm]{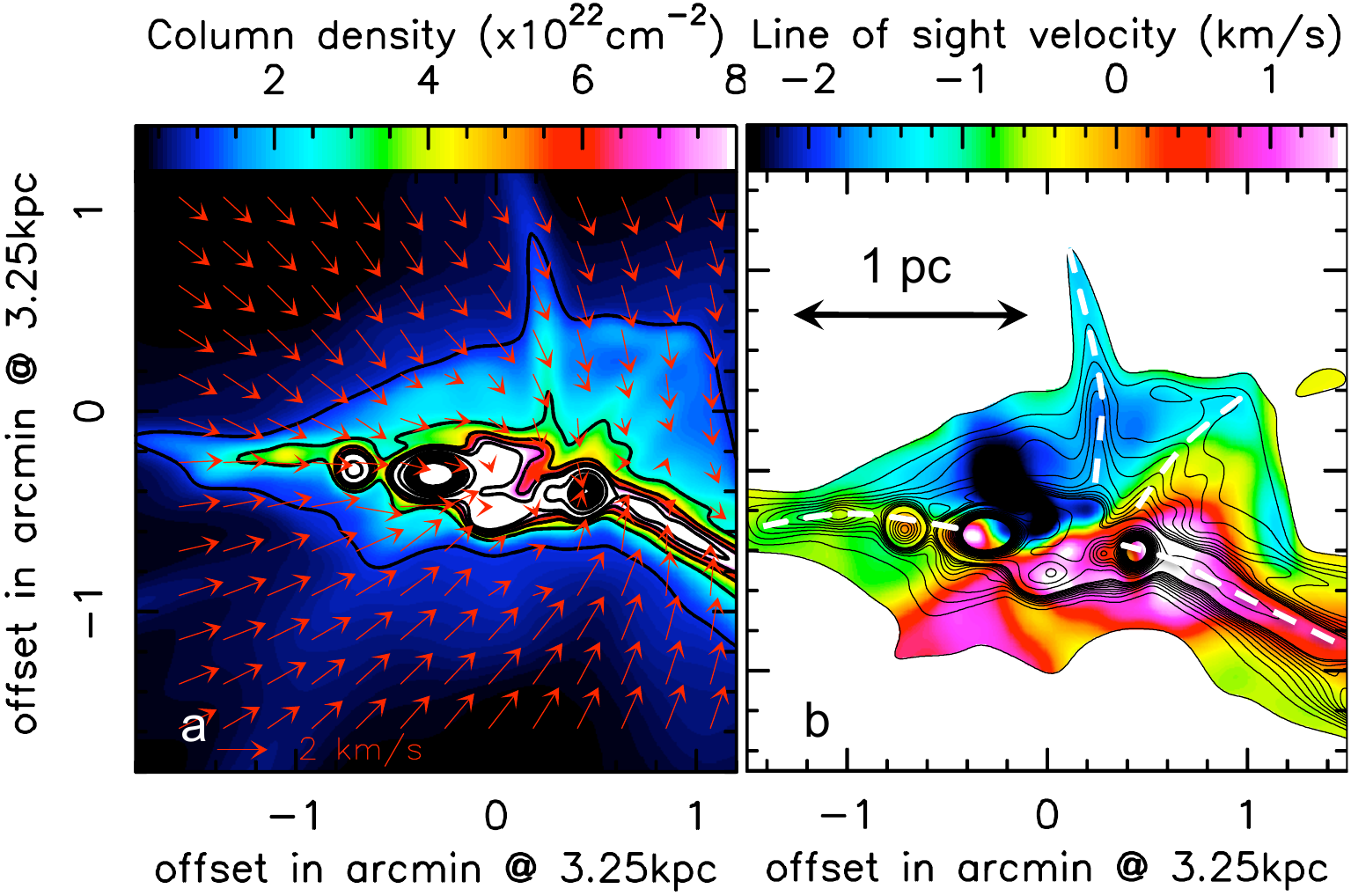}
 \caption{ Snapshot of a MHD simulation of a 10,000 M$_{\odot}$ collapsing cloud \citep[see Appendix C for more details; ][]{schneider2010}. (a) Column density (colour and contours) smoothed to the resolution of the ALMA data (5\arcsec). The arrows show the plane-of-the-sky velocity field. We see that gas flows along filaments and also off the filaments. (b) Dense gas line-of-sight velocity field (colour scale) smoothed to the resolution of the ALMA data. We highlighted the filaments by white dashed lines. The contours are the same as in panel (a).}
              \label{velocity}%
    \end{figure}

To illustrate some of the expected signatures of globally collapsing clouds we
present, in Fig.~\ref{velocity} a snapshot of a published MHD simulation modelling the evolution of a
turbulent and magnetized 10\,000 M$_{\odot}$ cloud, that was initially designed to reproduce some of the
observational signatures of the DR21 region \citep[][ see Appendix C for more details on the simulation]{schneider2010}. 
Overall, this simulation shows some similarities with SDC335, i.e. massive cores in the centre, the formation of filaments converging
towards these cores, and a velocity field resembling the one observed in SDC335 (see Fig.~\ref{n2hp}c). But most importantly,  Fig.~\ref{velocity} shows that although a fraction of the gas is indeed collapsing along the filaments, a large portion is collapsing off filaments. In this case the filamentary accretion observed along the filaments  represents only the tip of the entire accretion towards the cloud centre. 

In the context of a global collapse scenario, the observed velocity field along the filaments is the consequence of the inflowing cold gas. We can therefore estimate the current infall rate of gas running through the filaments using {\bf $\dot{M}_{inf}=N_{fil} \pi R_{fil}^2V_{inf}\rho_{fil}$}, where $N_{fil}$ is the number of filaments,  $R_{fil}$ is the filament cross-section radius, $V_{inf}$ is the gas infall velocity, and $\rho_{fil}$ is the gas volume density. With six filaments,  an infall velocity of  $0.7(\pm0.2)$~km/s, a filament section radius of 0.15~pc, and a density of $4(\pm1)\times10^4$~cm$^{-3}$, we derive an infall rate of   $0.7 (\pm0.3)\times10^{-3}$~M$_{\odot}$/yr. At this rate, a total mass of  $210(\pm90)$~M$_{\odot}$ would have been gathered in the centre by filamentary accretion within a free-fall time $t_{ff}\sim3\times10^5$~yr. This is slightly less than half of the cumulated core masses. However, less than 20\% of the SDC335 mass is lying within the filaments (cf Section 3). Assuming that the remaining gas is collapsing off filaments at a similar infall velocity, as observed in the simulations (see Fig.~\ref{velocity}a), the total accretion rate becomes  $\dot{M}_{inf}=4\pi R_{sph}^2 V_{inf}\rho_{sph}$, where R$_{sph}$ is the radius of the considered spherical volume and $\rho_{sph}$ is the density at that radius. At the radius of the Centre region, $R_{sph}=0.6$~pc and  $\rho_{sph}=1.3(\pm0.2)\times10^4$~cm$^{-3}$, which leads to $\dot{M}_{inf}=2.5(\pm1.0)\times10^{-3}$~M$_{\odot}$/yr. With this accretion rate $750(\pm300)$~M$_{\odot}$ of pristine gas is trapped inside the Centre region every cloud free-fall time. This is enough to  double the mass of material currently present in the Centre region in $3.5^{+2.2}_{-1.0}$ cloud free-fall times. This is few free-fall times longer than the typical timescale  over which simulations modelling the evolution of massive star-forming clouds evolve \citep[e.g.][]{smith2009,krumholz2012}. However, free-fall times in simulations are estimated at the initial average density $\bar{\rho}_{ini}$ while here $t_{ff}$ is calculated from the SDC335 current average density $\bar{\rho}_{cur}$. Therefore, a fair comparison with models  would imply that we estimate $t_{ff}$ from SDC335 $\bar{\rho}_{ini}$, which would then increase $t_{ff}$ as $(\bar{\rho}_{cur}/\bar{\rho}_{ini})^{1/2}$. Altogether, evidence indicates that a significant fraction, if not all, of the SDC335 core mass could have been built through the parsec-scale collapse of their parental cloud over a few cloud free-fall times.

\section{Summary and conclusion}

SDC335 is a massive ($5500\pm800$~M$_{\odot}$) IRDC with two massive star-forming cores located in its centre, one of which is likely to be an early O-type star progenitor. This core has an estimated mass of $545^{+770}_{-385}$~M$_{\odot}$ in a deconvolved diameter of $\sim0.05$~pc, which makes it one of the most massive protostellar cores ever observed in the Galaxy. A theoretical argument based on volume-density PDFs of molecular clouds suggests that such a concentration of mass must occur through the large-scale collapse of the surrounding cloud. This scenario was supported by several observational facts presented in this paper:  optically thick molecular line observations showing extended collapse signatures; a virial parameter distinctly lower than 1; a velocity field consistent with models of globally collapsing molecular clouds; accretion rates that are high enough to provide an additional $750(\pm300)$~M$_{\odot}$ of pristine gas to the central pc-size region of SDC335 per cloud free-fall time.  Altogether, these observations strongly suggest that the SDC335 massive star-forming cores  managed to build-up their large masses as a result of the supersonic global collapse of their surrounding cloud. 

It is always adventurous to draw general conclusions based on a single example. However, there are several sources now for which large-scale infall has been suggested to play a major role in building up the mass of star-forming cores \citep[e.g. ][]{peretto2006,schneider2010,barnes2010,galvanmadrid2010,liu2012}. Global infall also naturally explains the mass segregation observed in SDC335 and other regions. Although a number of questions remain to be answered, it is becoming clear that large-scale evolution of molecular clouds is the key to the mass determination process of massive stars.

\begin{acknowledgements}
 We thank the anonymous referee for his thorough report which significantly improved the quality of this paper. NP was supported by a CEA/Marie Curie Eurotalents fellowship and benefited from the support of the European Research Council advanced grant ORISTARS (Grant Agreement no. 291294). ADC was supported by the PROBeS project funded by the French National Research Agency (ANR). JEP has received funding from the European Community Seventh Framework Programme (/FP7/2007-2013/) under grant agreement No. 229517. NS was supported by the  ANR-11-BS56-010 STARFICH project. We also acknowledge the support of the European ALMA Regional Centre (ARC) and the UK ARC node. This paper makes use of the following ALMA data: ADS/JAO.ALMA\#2011.0.00474.S. ALMA is a partnership of ESO (representing its member states), NFS (USA) and NINS (Japan), together with NRC (Canada) and NSC and ASIAA (Taiwan), in cooperation with the Republic of Chile. The Joint ALMA Observatory is operated by ESO, AUI/NRAO and NAOJ. The Mopra radio telescope is part of the Australia Telescope National Facility, which is funded by the Commonwealth of Australia for operation as a National Facility managed by CSIRO. The University of New South Wales Digital Filter Bank used for the observations with the Mopra telescope was provided with the support from Australian Research Council. 
   \end{acknowledgements}

\bibliographystyle{aa}
\bibliography{references}

\begin{thebibliography}{78}
\expandafter\ifx\csname natexlab\endcsname\relax\def\natexlab#1{#1}\fi

\bibitem[{{Andr{\'e}} {et~al.}(2010){Andr{\'e}}, {Men'shchikov}, {Bontemps},
  {K{\"o}nyves}, {Motte}, {}, {Didelon}, {Minier}, {Saraceno}, {Ward-Thompson},
  {di Francesco}, {White}, {Molinari}, {Testi}, {Abergel}, {Griffin},
  {Henning}, {Royer}, {Mer{\'{\i}}n}, {Vavrek}, {Attard}, {Arzoumanian},
  {Wilson}, {Ade}, {Aussel}, {Baluteau}, {Benedettini}, {Bernard}, {Blommaert},
  {Cambr{\'e}sy}, {Cox}, {di Giorgio}, {Hargrave}, {Hennemann}, {Huang},
  {Kirk}, {Krause}, {Launhardt}, {Leeks}, {Le Pennec}, {Li}, {Martin}, {Maury},
  {Olofsson}, {Omont}, {Peretto}, {Pezzuto}, {Prusti}, {Roussel}, {Russeil},
  {Sauvage}, {Sibthorpe}, {Sicilia-Aguilar}, {Spinoglio}, {Waelkens},
  {Woodcraft}, \& {Zavagno}}]{andre2010}
{Andr{\'e}}, P., {Men'shchikov}, A., {Bontemps}, S., {et~al.} 2010, \aap, 518,
  L102+

\bibitem[{{Barnes} {et~al.}(2010){Barnes}, {Yonekura}, {Ryder}, {Hopkins},
  {Miyamoto}, {Furukawa}, \& {Fukui}}]{barnes2010}
{Barnes}, P.~J., {Yonekura}, Y., {Ryder}, S.~D., {et~al.} 2010, \mnras, 402, 73

\bibitem[{{Beckwith} {et~al.}(1990){Beckwith}, {Sargent}, {Chini}, \&
  {Guesten}}]{beckwith1990}
{Beckwith}, S.~V.~W., {Sargent}, A.~I., {Chini}, R.~S., \& {Guesten}, R. 1990,
  \aj, 99, 924

\bibitem[{{Bertoldi} \& {McKee}(1992)}]{bertoldi1992}
{Bertoldi}, F. \& {McKee}, C.~F. 1992, \apj, 395, 140

\bibitem[{{Beuther} {et~al.}(2007){Beuther}, {Churchwell}, {McKee}, \&
  {Tan}}]{beuther2007b}
{Beuther}, H., {Churchwell}, E.~B., {McKee}, C.~F., \& {Tan}, J.~C. 2007,
  Protostars and Planets V, 165

\bibitem[{{Beuther} \& {Schilke}(2004)}]{beuther2004}
{Beuther}, H. \& {Schilke}, P. 2004, Science, 303, 1167

\bibitem[{{Beuther} {et~al.}(2002){Beuther}, {Schilke}, {Gueth}, {McCaughrean},
  {Andersen}, {Sridharan}, \& {Menten}}]{beuther2002}
{Beuther}, H., {Schilke}, P., {Gueth}, F., {et~al.} 2002, \aap, 387, 931

\bibitem[{{Beuther} {et~al.}(2003){Beuther}, {Schilke}, \&
  {Stanke}}]{beuther2003}
{Beuther}, H., {Schilke}, P., \& {Stanke}, T. 2003, \aap, 408, 601

\bibitem[{{Beuther} {et~al.}(2005){Beuther}, {Sridharan}, \&
  {Saito}}]{beuther2005}
{Beuther}, H., {Sridharan}, T.~K., \& {Saito}, M. 2005, \apjl, 634, L185

\bibitem[{{Beuther} {et~al.}(2012){Beuther}, {Tackenberg}, {Linz}, {Henning},
  {Krause}, {Ragan}, {Nielbock}, {Launhardt}, {Schmiedeke}, {Schuller},
  {Carlhoff}, {Nguyen-Luong}, \& {Sakai}}]{beuther2012}
{Beuther}, H., {Tackenberg}, J., {Linz}, H., {et~al.} 2012, \aap, 538, A11

\bibitem[{{Bonnell} {et~al.}(2004){Bonnell}, {Vine}, \& {Bate}}]{bonnell2004}
{Bonnell}, I.~A., {Vine}, S.~G., \& {Bate}, M.~R. 2004, \mnras, 349, 735

\bibitem[{{Bontemps} {et~al.}(2010){Bontemps}, {Motte}, {Csengeri}, \&
  {Schneider}}]{bontemps2010}
{Bontemps}, S., {Motte}, F., {Csengeri}, T., \& {Schneider}, N. 2010, \aap,
  524, A18

\bibitem[{{Busquet} {et~al.}(2011){Busquet}, {Estalella}, {Zhang}, {Viti},
  {Palau}, {Ho}, \& {S{\'a}nchez-Monge}}]{busquet2011}
{Busquet}, G., {Estalella}, R., {Zhang}, Q., {et~al.} 2011, \aap, 525, A141

\bibitem[{{Butler} \& {Tan}(2009)}]{butler2009}
{Butler}, M.~J. \& {Tan}, J.~C. 2009, \apj, 696, 484

\bibitem[{{Caswell} {et~al.}(2011){Caswell}, {Fuller}, {Green}, {Avison},
  {Breen}, {Ellingsen}, {Gray}, {Pestalozzi}, {Quinn}, {Thompson}, \&
  {Voronkov}}]{caswell2011}
{Caswell}, J.~L., {Fuller}, G.~A., {Green}, J.~A., {et~al.} 2011, \mnras, 417,
  1964

\bibitem[{{Chabrier}(2003)}]{chabrier2003}
{Chabrier}, G. 2003, in IAU Symposium, Vol. 221, IAU Symposium, 67P--+

\bibitem[{{Churchwell} {et~al.}(2009){Churchwell}, {Babler}, {Meade},
  {Whitney}, {Benjamin}, {Indebetouw}, {Cyganowski}, {Robitaille}, {Povich},
  {Watson}, \& {Bracker}}]{churchwell2009}
{Churchwell}, E., {Babler}, B.~L., {Meade}, M.~R., {et~al.} 2009, \pasp, 121,
  213

\bibitem[{{Crutcher}(2012)}]{crutcher2012}
{Crutcher}, R.~M. 2012, \araa, 50, 29

\bibitem[{{Csengeri} {et~al.}(2011){Csengeri}, {Bontemps}, {Schneider},
  {Motte}, {Gueth}, \& {Hora}}]{csengeri2011}
{Csengeri}, T., {Bontemps}, S., {Schneider}, N., {et~al.} 2011, \apjl, 740, L5

\bibitem[{{Cyganowski} {et~al.}(2011){Cyganowski}, {Brogan}, {Hunter}, \&
  {Churchwell}}]{cyganowski2011}
{Cyganowski}, C.~J., {Brogan}, C.~L., {Hunter}, T.~R., \& {Churchwell}, E.
  2011, \apj, 743, 56

\bibitem[{{Cyganowski} {et~al.}(2008){Cyganowski}, {Whitney}, {Holden},
  {Braden}, {Brogan}, {Churchwell}, {Indebetouw}, {Watson}, {Babler},
  {Benjamin}, {Gomez}, {Meade}, {Povich}, {Robitaille}, \&
  {Watson}}]{cyganowski2008}
{Cyganowski}, C.~J., {Whitney}, B.~A., {Holden}, E., {et~al.} 2008, \aj, 136,
  2391

\bibitem[{{De Vries} \& {Myers}(2005)}]{devries2005}
{De Vries}, C.~H. \& {Myers}, P.~C. 2005, \apj, 620, 800

\bibitem[{{Duarte-Cabral} {et~al.}(2013){Duarte-Cabral}, {Bontemps}, {Motte},
  {Hennemann}, {Schneider}, \& {Andr\'e}}]{duartecabral2013}
{Duarte-Cabral}, A., {Bontemps}, S., {Motte}, F., {et~al.} 2013, submitted to
  \aap

\bibitem[{{Fuller} {et~al.}(2005){Fuller}, {Williams}, \&
  {Sridharan}}]{Fuller2005}
{Fuller}, G.~A., {Williams}, S.~J., \& {Sridharan}, T.~K. 2005, \aap, 442, 949

\bibitem[{{Galv{\'a}n-Madrid} {et~al.}(2010){Galv{\'a}n-Madrid}, {Zhang},
  {Keto}, {Ho}, {Zapata}, {Rodr{\'{\i}}guez}, {Pineda}, \&
  {V{\'a}zquez-Semadeni}}]{galvanmadrid2010}
{Galv{\'a}n-Madrid}, R., {Zhang}, Q., {Keto}, E., {et~al.} 2010, \apj, 725, 17

\bibitem[{{Garay} {et~al.}(2002){Garay}, {Brooks}, {Mardones}, {Norris}, \&
  {Burton}}]{garay2002}
{Garay}, G., {Brooks}, K.~J., {Mardones}, D., {Norris}, R.~P., \& {Burton},
  M.~G. 2002, \apj, 579, 678

\bibitem[{{Griffin} {et~al.}(2010){Griffin}, {Abergel}, {Abreu}, {Ade},
  {Andr{\'e}}, {Augueres}, {Babbedge}, {Bae}, {Baillie}, {Baluteau}, {Barlow},
  {Bendo}, {Benielli}, {Bock}, {Bonhomme}, {Brisbin}, {Brockley-Blatt},
  {Caldwell}, {Cara}, {Castro-Rodriguez}, {Cerulli}, {Chanial}, {Chen},
  {Clark}, {Clements}, {Clerc}, {Coker}, {Communal}, {Conversi}, {Cox},
  {Crumb}, {Cunningham}, {Daly}, {Davis}, {de Antoni}, {Delderfield}, {Devin},
  {di Giorgio}, {Didschuns}, {Dohlen}, {Donati}, {Dowell}, {Dowell}, {Duband},
  {Dumaye}, {Emery}, {Ferlet}, {Ferrand}, {Fontignie}, {Fox}, {Franceschini},
  {Frerking}, {Fulton}, {Garcia}, {Gastaud}, {Gear}, {Glenn}, {Goizel},
  {Griffin}, {Grundy}, {Guest}, {Guillemet}, {Hargrave}, {Harwit}, {Hastings},
  {Hatziminaoglou}, {Herman}, {Hinde}, {Hristov}, {Huang}, {Imhof}, {Isaak},
  {Israelsson}, {Ivison}, {Jennings}, {Kiernan}, {King}, {Lange}, {Latter},
  {Laurent}, {Laurent}, {Leeks}, {Lellouch}, {Levenson}, {Li}, {Li},
  {Lilienthal}, {Lim}, {Liu}, {Lu}, {Madden}, {Mainetti}, {Marliani}, {McKay},
  {Mercier}, {Molinari}, {Morris}, {Moseley}, {Mulder}, {Mur}, {Naylor},
  {Nguyen}, {O'Halloran}, {Oliver}, {Olofsson}, {Olofsson}, {Orfei}, {Page},
  {Pain}, {Panuzzo}, {Papageorgiou}, {Parks}, {Parr-Burman}, {Pearce},
  {Pearson}, {P{\'e}rez-Fournon}, {Pinsard}, {Pisano}, {Podosek}, {Pohlen},
  {Polehampton}, {Pouliquen}, {Rigopoulou}, {Rizzo}, {Roseboom}, {Roussel},
  {Rowan-Robinson}, {Rownd}, {Saraceno}, {Sauvage}, {Savage}, {Savini},
  {Sawyer}, {Scharmberg}, {Schmitt}, {}, {Schulz}, {Schwartz}, {Shafer},
  {Shupe}, {Sibthorpe}, {Sidher}, {Smith}, {Smith}, {Smith}, {Spencer},
  {Stobie}, {Sudiwala}, {Sukhatme}, {Surace}, {Stevens}, {Swinyard}, {Trichas},
  {Tourette}, {Triou}, {Tseng}, {Tucker}, {Turner}, {Vaccari}, {Valtchanov},
  {Vigroux}, {Virique}, {Voellmer}, {Walker}, {Ward}, {Waskett}, {Weilert},
  {Wesson}, {White}, {Whitehouse}, {Wilson}, {Winter}, {Woodcraft}, {Wright},
  {Xu}, {Zavagno}, {Zemcov}, {Zhang}, \& {Zonca}}]{griffin2010}
{Griffin}, M.~J., {Abergel}, A., {Abreu}, A., {et~al.} 2010, \aap, 518, L3+

\bibitem[{{Hennebelle} \& {Chabrier}(2008)}]{hennebelle2008}
{Hennebelle}, P. \& {Chabrier}, G. 2008, \apj, 684, 395

\bibitem[{{Hennemann} {et~al.}(2012){Hennemann}, {Motte}, {}, {Didelon},
  {Hill}, {Arzoumanian}, {Bontemps}, {Csengeri}, {Andr{\'e}}, {Konyves},
  {Louvet}, {Marston}, {Men'shchikov}, {Minier}, {Nguyen Luong}, {Palmeirim},
  {Peretto}, {Sauvage}, {Zavagno}, {Anderson}, {Bernard}, {Di Francesco},
  {Elia}, {Li}, {Martin}, {Molinari}, {Pezzuto}, {Russeil}, {Rygl}, {Schisano},
  {Spinoglio}, {Sousbie}, {Ward-Thompson}, \& {White}}]{hennemann2012}
{Hennemann}, M., {Motte}, F., {}, N., {et~al.} 2012, \aap, 543, L3

\bibitem[{{Hildebrand}(1983)}]{hildebrand1983}
{Hildebrand}, R.~H. 1983, \qjras, 24, 267

\bibitem[{{Hoare}(2005)}]{hoare2005}
{Hoare}, M.~G. 2005, \apss, 295, 203

\bibitem[{{Hogerheijde} \& {van der Tak}(2000)}]{hogerheijde2000}
{Hogerheijde}, M.~R. \& {van der Tak}, F.~F.~S. 2000, \aap, 362, 697

\bibitem[{{Kauffmann} \& {Pillai}(2010)}]{kauffmann2010}
{Kauffmann}, J. \& {Pillai}, T. 2010, \apjl, 723, L7

\bibitem[{{Kroupa}(2002)}]{kroupa2002}
{Kroupa}, P. 2002, Science, 295, 82

\bibitem[{{Krumholz} {et~al.}(2012){Krumholz}, {Klein}, \&
  {McKee}}]{krumholz2012}
{Krumholz}, M.~R., {Klein}, R.~I., \& {McKee}, C.~F. 2012, \apj, 754, 71

\bibitem[{{Lada} \& {Lada}(2003)}]{lada2003}
{Lada}, C.~J. \& {Lada}, E.~A. 2003, \araa, 41, 57

\bibitem[{{Ladd} {et~al.}(2005){Ladd}, {Purcell}, {Wong}, \&
  {Robertson}}]{ladd2005}
{Ladd}, N., {Purcell}, C., {Wong}, T., \& {Robertson}, S. 2005, \pasa, 22, 62

\bibitem[{{Langer} \& {Penzias}(1993)}]{langer1993}
{Langer}, W.~D. \& {Penzias}, A.~A. 1993, \apj, 408, 539

\bibitem[{{Liu} {et~al.}(2012){Liu}, {Jim{\'e}nez-Serra}, {Ho}, {Chen},
  {Zhang}, \& {Li}}]{liu2012}
{Liu}, H.~B., {Jim{\'e}nez-Serra}, I., {Ho}, P.~T.~P., {et~al.} 2012, \apj,
  756, 10

\bibitem[{{McKee} \& {Tan}(2003)}]{mckee2003}
{McKee}, C.~F. \& {Tan}, J.~C. 2003, \apj, 585, 850

\bibitem[{{McMullin} {et~al.}(2007){McMullin}, {Waters}, {Schiebel}, {Young},
  \& {Golap}}]{mcmullin2007}
{McMullin}, J.~P., {Waters}, B., {Schiebel}, D., {Young}, W., \& {Golap}, K.
  2007, in Astronomical Society of the Pacific Conference Series, Vol. 376,
  Astronomical Data Analysis Software and Systems XVI, ed. R.~A. {Shaw},
  F.~{Hill}, \& D.~J. {Bell}, 127

\bibitem[{{Molinari} {et~al.}(2010){Molinari}, {Swinyard}, {Bally}, {Barlow},
  {Bernard}, {Martin}, {Moore}, {Noriega-Crespo}, {Plume}, {Testi}, {Zavagno},
  {Abergel}, {Ali}, {Anderson}, {Andr{\'e}}, {Baluteau}, {Battersby},
  {Beltr{\'a}n}, {Benedettini}, {Billot}, {Blommaert}, {Bontemps}, {Boulanger},
  {Brand}, {Brunt}, {Burton}, {Calzoletti}, {Carey}, {Caselli}, {Cesaroni},
  {Cernicharo}, {Chakrabarti}, {Chrysostomou}, {Cohen}, {Compiegne}, {de
  Bernardis}, {de Gasperis}, {di Giorgio}, {Elia}, {Faustini}, {Flagey},
  {Fukui}, {Fuller}, {Ganga}, {Garcia-Lario}, {Glenn}, {Goldsmith}, {Griffin},
  {Hoare}, {Huang}, {Ikhenaode}, {Joblin}, {Joncas}, {Juvela}, {Kirk},
  {Lagache}, {Li}, {Lim}, {Lord}, {Marengo}, {Marshall}, {Masi}, {Massi},
  {Matsuura}, {Minier}, {Miville-Desch{\^e}nes}, {Montier}, {Morgan}, {Motte},
  {Mottram}, {M{\"u}ller}, {Natoli}, {Neves}, {Olmi}, {Paladini}, {Paradis},
  {Parsons}, {Peretto}, {Pestalozzi}, {Pezzuto}, {Piacentini}, {Piazzo},
  {Polychroni}, {Pomar{\`e}s}, {Popescu}, {Reach}, {Ristorcelli}, {Robitaille},
  {Robitaille}, {Rod{\'o}n}, {Roy}, {Royer}, {Russeil}, {Saraceno}, {Sauvage},
  {Schilke}, {Schisano}, {}, {Schuller}, {Schulz}, {Sibthorpe}, {Smith},
  {Smith}, {Spinoglio}, {Stamatellos}, {Strafella}, {Stringfellow}, {Sturm},
  {Taylor}, {Thompson}, {Traficante}, {Tuffs}, {Umana}, {Valenziano}, {Vavrek},
  {Veneziani}, {Viti}, {Waelkens}, {Ward-Thompson}, {White}, {Wilcock},
  {Wyrowski}, {Yorke}, \& {Zhang}}]{molinari2010}
{Molinari}, S., {Swinyard}, B., {Bally}, J., {et~al.} 2010, \aap, 518, L100+

\bibitem[{{Molinari} {et~al.}(1998){Molinari}, {Testi}, {Brand}, {Cesaroni}, \&
  {Palla}}]{molinari1998}
{Molinari}, S., {Testi}, L., {Brand}, J., {Cesaroni}, R., \& {Palla}, F. 1998,
  \apjl, 505, L39

\bibitem[{{Myers}(2009)}]{myers2009}
{Myers}, P.~C. 2009, \apj, 700, 1609

\bibitem[{{Myers} {et~al.}(1996){Myers}, {Mardones}, {Tafalla}, {Williams}, \&
  {Wilner}}]{myers1996}
{Myers}, P.~C., {Mardones}, D., {Tafalla}, M., {Williams}, J.~P., \& {Wilner},
  D.~J. 1996, \apjl, 465, L133

\bibitem[{{Padoan} {et~al.}(1997){Padoan}, {Nordlund}, \& {Jones}}]{padoan1997}
{Padoan}, P., {Nordlund}, A., \& {Jones}, B.~J.~T. 1997, \mnras, 288, 145

\bibitem[{{Paradis} {et~al.}(2010){Paradis}, {Veneziani}, {Noriega-Crespo},
  {Paladini}, {Piacentini}, {Bernard}, {de Bernardis}, {Calzoletti},
  {Faustini}, {Martin}, {Masi}, {Montier}, {Natoli}, {Ristorcelli}, {Thompson},
  {Traficante}, \& {Molinari}}]{paradis2010}
{Paradis}, D., {Veneziani}, M., {Noriega-Crespo}, A., {et~al.} 2010, \aap, 520,
  L8

\bibitem[{{Peretto} {et~al.}(2006){Peretto}, {Andr{\'e}}, \&
  {Belloche}}]{peretto2006}
{Peretto}, N., {Andr{\'e}}, P., \& {Belloche}, A. 2006, \aap, 445, 979

\bibitem[{{Peretto} {et~al.}(2012){Peretto}, {Andr{\'e}}, {K{\"o}nyves}, {},
  {Arzoumanian}, {Palmeirim}, {Didelon}, {Attard}, {Bernard}, {Di Francesco},
  {Elia}, {Hennemann}, {Hill}, {Kirk}, {Men'shchikov}, {Motte}, {Nguyen Luong},
  {Roussel}, {Sousbie}, {Testi}, {Ward-Thompson}, {White}, \&
  {Zavagno}}]{peretto2012}
{Peretto}, N., {Andr{\'e}}, P., {K{\"o}nyves}, V., {et~al.} 2012, \aap, 541,
  A63

\bibitem[{{Peretto} \& {Fuller}(2009)}]{peretto2009}
{Peretto}, N. \& {Fuller}, G.~A. 2009, \aap, 505, 405

\bibitem[{{Peretto} \& {Fuller}(2010)}]{peretto2010a}
{Peretto}, N. \& {Fuller}, G.~A. 2010, \apj, 723, 555

\bibitem[{{Peretto} {et~al.}(2010){Peretto}, {Fuller}, {Plume}, {Anderson},
  {Bally}, {Battersby}, {Beltran}, {Bernard}, {Calzoletti}, {Digiorgio},
  {Faustini}, {Kirk}, {Lenfestey}, {Marshall}, {Martin}, {Molinari}, {Montier},
  {Motte}, {Ristorcelli}, {Rod{\'o}n}, {Smith}, {Traficante}, {Veneziani},
  {Ward-Thompson}, \& {Wilcock}}]{peretto2010b}
{Peretto}, N., {Fuller}, G.~A., {Plume}, R., {et~al.} 2010, \aap, 518, L98+

\bibitem[{{Peretto} {et~al.}(2007){Peretto}, {Hennebelle}, \&
  {Andr{\'e}}}]{peretto2007}
{Peretto}, N., {Hennebelle}, P., \& {Andr{\'e}}, P. 2007, \aap, 464, 983

\bibitem[{{Pilbratt} {et~al.}(2010){Pilbratt}, {Riedinger}, {Passvogel},
  {Crone}, {Doyle}, {Gageur}, {Heras}, {Jewell}, {Metcalfe}, {Ott}, \&
  {Schmidt}}]{pilbratt2010}
{Pilbratt}, G.~L., {Riedinger}, J.~R., {Passvogel}, T., {et~al.} 2010, \aap,
  518, L1+

\bibitem[{{Pillai} {et~al.}(2011){Pillai}, {Kauffmann}, {Wyrowski}, {Hatchell},
  {Gibb}, \& {Thompson}}]{pillai2011}
{Pillai}, T., {Kauffmann}, J., {Wyrowski}, F., {et~al.} 2011, \aap, 530, A118

\bibitem[{{Poglitsch} {et~al.}(2010){Poglitsch}, {Waelkens}, {Geis},
  {Feuchtgruber}, {Vandenbussche}, {Rodriguez}, {Krause}, {Renotte}, {van
  Hoof}, {Saraceno}, {Cepa}, {Kerschbaum}, {Agn{\`e}se}, {Ali}, {Altieri},
  {Andreani}, {Augueres}, {Balog}, {Barl}, {Bauer}, {Belbachir}, {Benedettini},
  {Billot}, {Boulade}, {Bischof}, {Blommaert}, {Callut}, {Cara}, {Cerulli},
  {Cesarsky}, {Contursi}, {Creten}, {De Meester}, {Doublier}, {Doumayrou},
  {Duband}, {Exter}, {Genzel}, {Gillis}, {Gr{\"o}zinger}, {Henning},
  {Herreros}, {Huygen}, {Inguscio}, {Jakob}, {Jamar}, {Jean}, {de Jong},
  {Katterloher}, {Kiss}, {Klaas}, {Lemke}, {Lutz}, {Madden}, {Marquet},
  {Martignac}, {Mazy}, {Merken}, {Montfort}, {Morbidelli}, {M{\"u}ller},
  {Nielbock}, {Okumura}, {Orfei}, {Ottensamer}, {Pezzuto}, {Popesso},
  {Putzeys}, {Regibo}, {Reveret}, {Royer}, {Sauvage}, {Schreiber}, {Stegmaier},
  {Schmitt}, {Schubert}, {Sturm}, {Thiel}, {Tofani}, {Vavrek}, {Wetzstein},
  {Wieprecht}, \& {Wiezorrek}}]{poglitsch2010}
{Poglitsch}, A., {Waelkens}, C., {Geis}, N., {et~al.} 2010, \aap, 518, L2+

\bibitem[{{Ragan} {et~al.}(2012){Ragan}, {Heitsch}, {Bergin}, \&
  {Wilner}}]{ragan2012}
{Ragan}, S.~E., {Heitsch}, F., {Bergin}, E.~A., \& {Wilner}, D. 2012, \apj,
  746, 174

\bibitem[{{Rathborne} {et~al.}(2011){Rathborne}, {Garay}, {Jackson},
  {Longmore}, {Zhang}, \& {Simon}}]{rathborne2011}
{Rathborne}, J.~M., {Garay}, G., {Jackson}, J.~M., {et~al.} 2011, \apj, 741,
  120

\bibitem[{{Rathborne} {et~al.}(2006){Rathborne}, {Jackson}, \&
  {Simon}}]{rathborne2006}
{Rathborne}, J.~M., {Jackson}, J.~M., \& {Simon}, R. 2006, \apj, 641, 389

\bibitem[{{Rathborne} {et~al.}(2008){Rathborne}, {Jackson}, {Zhang}, \&
  {Simon}}]{rathborne2008}
{Rathborne}, J.~M., {Jackson}, J.~M., {Zhang}, Q., \& {Simon}, R. 2008, \apj,
  689, 1141

\bibitem[{{Rathborne} {et~al.}(2007){Rathborne}, {Simon}, \&
  {Jackson}}]{rathborne2007}
{Rathborne}, J.~M., {Simon}, R., \& {Jackson}, J.~M. 2007, \apj, 662, 1082

\bibitem[{{Reid} {et~al.}(2009){Reid}, {Menten}, {Zheng}, {Brunthaler},
  {Moscadelli}, {Xu}, {Zhang}, {Sato}, {Honma}, {Hirota}, {Hachisuka}, {Choi},
  {Moellenbrock}, \& {Bartkiewicz}}]{reid2009}
{Reid}, M.~J., {Menten}, K.~M., {Zheng}, X.~W., {et~al.} 2009, \apj, 700, 137

\bibitem[{{Ren} {et~al.}(2012){Ren}, {Wu}, {Zhu}, {Liu}, {Peng}, {Qin}, \&
  {Li}}]{ren2012}
{Ren}, Z., {Wu}, Y., {Zhu}, M., {et~al.} 2012, \mnras, 422, 1098

\bibitem[{{Rod{\'o}n} {et~al.}(2012){Rod{\'o}n}, {Beuther}, \&
  {Schilke}}]{rodon2012}
{Rod{\'o}n}, J.~A., {Beuther}, H., \& {Schilke}, P. 2012, \aap, 545, A51

\bibitem[{{Savage} {et~al.}(2002){Savage}, {Apponi}, {Ziurys}, \&
  {Wyckoff}}]{savage2002}
{Savage}, C., {Apponi}, A.~J., {Ziurys}, L.~M., \& {Wyckoff}, S. 2002, \apj,
  578, 211

\bibitem[{{Schneider} {et~al.}(2010){Schneider}, {Csengeri}, {Bontemps},
  {Motte}, {Simon}, {Hennebelle}, {Federrath}, \& {Klessen}}]{schneider2010}
{Schneider}, N., {Csengeri}, T., {Bontemps}, S., {et~al.} 2010, \aap, 520, A49

\bibitem[{{Schneider} {et~al.}(2012){Schneider}, {Csengeri}, {Hennemann},
  {Motte}, {Didelon}, {Federrath}, {Bontemps}, {Di Francesco}, {Arzoumanian},
  {Minier}, {Andr{\'e}}, {Hill}, {Zavagno}, {Nguyen-Luong}, {Attard},
  {Bernard}, {Elia}, {Fallscheer}, {Griffin}, {Kirk}, {Klessen}, {K{\"o}nyves},
  {Martin}, {Men'shchikov}, {Palmeirim}, {Peretto}, {Pestalozzi}, {Russeil},
  {Sadavoy}, {Sousbie}, {Testi}, {Tremblin}, {Ward-Thompson}, \&
  {White}}]{schneider2012}
{Schneider}, N., {Csengeri}, T., {Hennemann}, M., {et~al.} 2012, \aap, 540, L11

\bibitem[{{Smith} {et~al.}(2009){Smith}, {Longmore}, \& {Bonnell}}]{smith2009}
{Smith}, R.~J., {Longmore}, S., \& {Bonnell}, I. 2009, \mnras, 400, 1775

\bibitem[{{Smith} {et~al.}(2012){Smith}, {Shetty}, {Stutz}, \&
  {Klessen}}]{smith2012}
{Smith}, R.~J., {Shetty}, R., {Stutz}, A.~M., \& {Klessen}, R.~S. 2012, \apj,
  750, 64

\bibitem[{{Traficante} {et~al.}(2011){Traficante}, {Calzoletti}, {Veneziani},
  {Ali}, {de Gasperis}, {di Giorgio}, {Faustini}, {Ikhenaode}, {Molinari},
  {Natoli}, {Pestalozzi}, {Pezzuto}, {Piacentini}, {Piazzo}, {Polenta}, \&
  {Schisano}}]{traficante2011}
{Traficante}, A., {Calzoletti}, L., {Veneziani}, M., {et~al.} 2011, \mnras,
  416, 2932

\bibitem[{{van der Tak} {et~al.}(2007){van der Tak}, {Black}, {Sch{\"o}ier},
  {Jansen}, \& {van Dishoeck}}]{vandertak2007}
{van der Tak}, F.~F.~S., {Black}, J.~H., {Sch{\"o}ier}, F.~L., {Jansen}, D.~J.,
  \& {van Dishoeck}, E.~F. 2007, \aap, 468, 627

\bibitem[{{Wang} {et~al.}(2011){Wang}, {Zhang}, {Wu}, \& {Zhang}}]{wang2011}
{Wang}, K., {Zhang}, Q., {Wu}, Y., \& {Zhang}, H. 2011, \apj, 735, 64

\bibitem[{{Wilcock} {et~al.}(2012){Wilcock}, {Ward-Thompson}, {Kirk},
  {Stamatellos}, {Whitworth}, {Battersby}, {Elia}, {Fuller}, {DiGiorgio},
  {Griffin}, {Molinari}, {Martin}, {Mottram}, {Peretto}, {Pestalozzi},
  {Schisano}, {Smith}, \& {Thompson}}]{wilcock2012}
{Wilcock}, L.~A., {Ward-Thompson}, D., {Kirk}, J.~M., {et~al.} 2012, \mnras,
  424, 716

\bibitem[{{Xu} {et~al.}(2008){Xu}, {Li}, {Hachisuka}, {Pandian}, {Menten}, \&
  {Henkel}}]{xu2008}
{Xu}, Y., {Li}, J.~J., {Hachisuka}, K., {et~al.} 2008, \aap, 485, 729

\bibitem[{{Zapata} {et~al.}(2009){Zapata}, {Ho}, {Schilke}, {Rodr{\'{\i}}guez},
  {Menten}, {Palau}, \& {Garrod}}]{zapata2009}
{Zapata}, L.~A., {Ho}, P.~T.~P., {Schilke}, P., {et~al.} 2009, \apj, 698, 1422

\bibitem[{{Zhang} \& {Wang}(2011)}]{zhang2011}
{Zhang}, Q. \& {Wang}, K. 2011, \apj, 733, 26

\bibitem[{{Zinchenko} {et~al.}(2009){Zinchenko}, {Caselli}, \&
  {Pirogov}}]{zinchenko2009}
{Zinchenko}, I., {Caselli}, P., \& {Pirogov}, L. 2009, \mnras, 395, 2234

\bibitem[{{Zinnecker} \& {Yorke}(2007)}]{zinnecker2007}
{Zinnecker}, H. \& {Yorke}, H.~W. 2007, \araa, 45, 481

\end{thebibliography}

\appendix

\section{Volume-density PDF calculations}

Volume-density PDFs of turbulent, non-self-gravitating molecular clouds can be described as a lognormal function of the logarithmic density contrast $\delta=\log(\rho/\bar{\rho})$ \citep{padoan1997,hennebelle2008}:

\begin{equation}
\mathcal{P}(\delta)=\frac{1}{\sqrt{2\pi\sigma_0^2}}\exp\left[-\frac{(\delta-\bar{\delta})^2}{2\sigma_0^2}\right] \, ,
\end{equation}
where $\sigma_0$ is the standard deviation of the distribution,  and $\bar{\delta}=-\sigma_0^2/2$. Furthermore, the standard deviation of this PDF can be written as $\sigma_0^2=\ln(1+b\mathcal{M}^2)$, where $\mathcal{M}$ is the Mach number and $b\simeq0.25$.

Integrating Equation (A.1), we can estimate the fraction of a cloud that lies above a certain density threshold, $\rho_{th}$, before gravity becomes important. Setting the two free parameters of Eq.~A.1 to the SDC335 observed values ($\bar{\rho}=1.3\times10^4$~cm$^{-3}$ and $\mathcal{M}=6$), we find that less than 0.01\% of the gas should lie above $\rho_{th}=1\times10^7$~cm$^{-3}$. This is more than three orders of magnitude   different from what is observed in SDC335.

Now we can estimate the density $\rho_{ini}$ (and $\delta_{ini}=\rho_{ini}/\bar{\rho}$) at which the following relation is fulfilled: 
\begin{equation}
\frac{M_{\rm{MM1}}}{M_{\rm{SDC335}}}=\int_{\delta_{ini}}^{\infty}{\mathcal{P}(\delta) d\delta} \, ,
\end{equation}
which is equivalent to
\begin{equation}
0.1=0.5\times(1-\rm{erf}[(\delta_{ini}-\bar{\delta})/\sqrt{2\sigma_0^2}]) \, ,
\end{equation}
This gives $\rho_{ini}=3.5\times10^4$~cm$^{-3}$. Then we can calculate the volume diameter in which the MM1 mass was initially contained,  using  $D_{ini}=2\times[3 M_{MM1}/(4\pi\rho_{ini})]^{1/3}=0.8$~pc. Note that here we used $\rho_{ini}$ as the mean density of the initial volume, while it formally is the minimum density within the volume under consideration. The true mean density is necessarily higher, although it cannot be too centrally concentrated either because this would not satisfy the volume density PDF. It is therefore reasonable to use $\rho_{ini}$ as the mean density, especially since the dependency of $D_{ini}$ on $\rho_{ini}$  is weak.

Alternatively, we can estimate the maximum $M_{ini}$  in the current MM1 volume that satisfies the volume density PDF:

\begin{equation}
\frac{M_{ini}}{M_{\rm{SDC335}}}=\int_{\delta_{ini}}^{\infty}{\mathcal{P}(\delta) d\delta} \, , 
\end{equation}
with $M_{ini}=\rho_{ini}\,V_{MM1} $. We find that $M_{ini} \simeq3$~M$_{\odot}$, which means that in this case nearly the whole MM1 mass must come from its surrounding. We can estimate the volume of this region by taking  $D_{ini}=2\times[3 M_{MM1}/(4\pi\bar{\rho})]^{1/3}=1.2$~pc.

\section{H$^{13}$CO$^+$(1-0) RATRAN modelling}

\begin{figure}
\centering
\includegraphics[width=8cm]{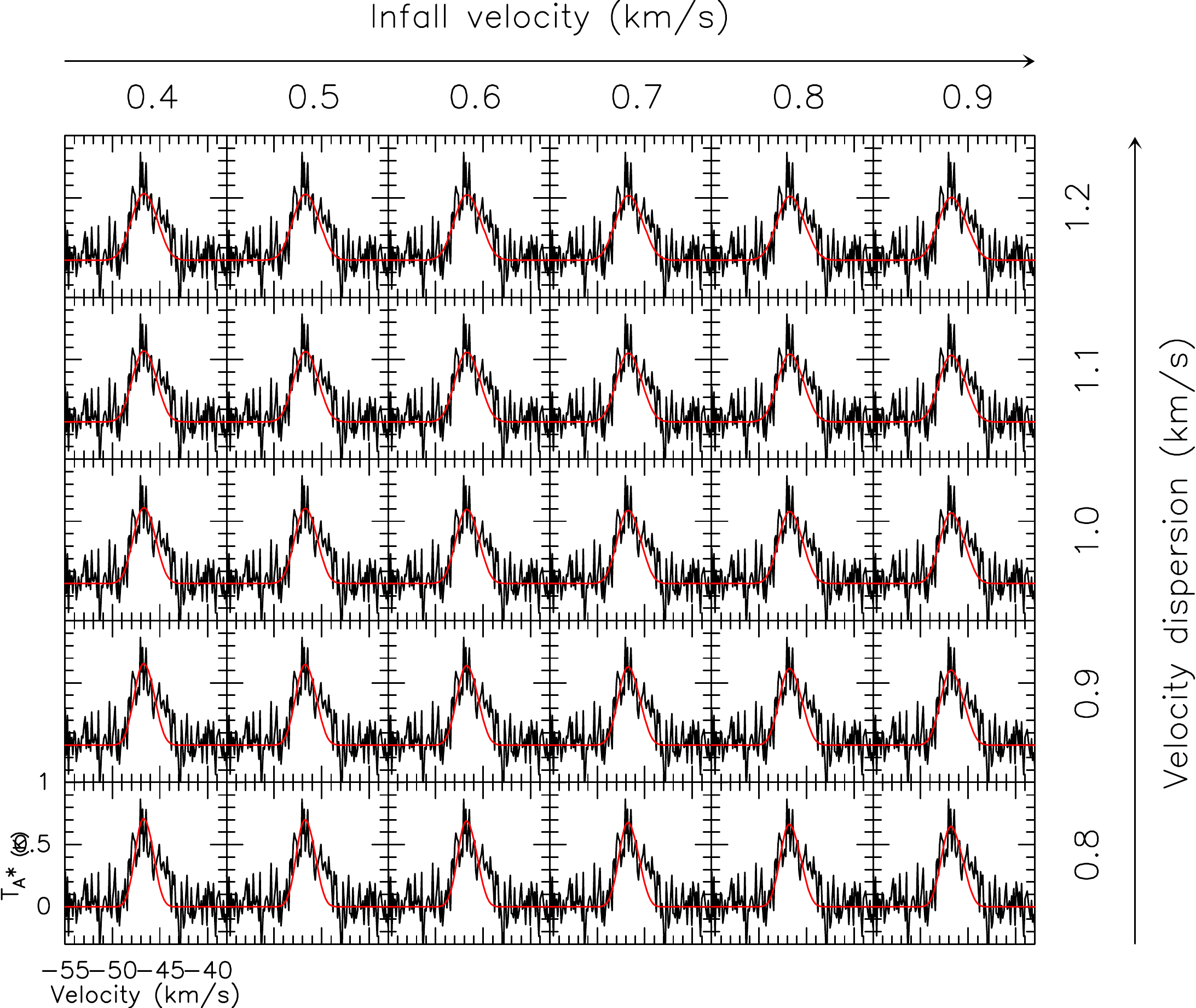}
 \caption{ Same as Fig.~\ref{specmodhcop}, but for the H$^{13}$CO$^+$(1-0) line.}
              \label{h13copmodel}%
    \end{figure}

 Figure~\ref{h13copmodel} presents the optically thin H$^{13}$CO$^+$(1-0) modelled spectra obtain with RATRAN for the cloud-collapse model discussed in Sec.~5.3. We see that the modelled lines overall match the observed spectrum quite well even though, in nearly all cases, the modelled spectrum is slightly too narrow. This can potentially be explained by a more complex infall profile than the one we used for these calculation, resulting in a slightly broader line. The modelled spectra have peak temperatures close to the observed one, which supports our choice of HCO$^+$ and H$^{13}$CO$^+$ abundances.

\section{Additional details on the MHD simulation}

The simulation presented in this study (Fig.~\ref{velocity} of the paper) was initially performed to model the DR21 region \citep{schneider2010}. It is an MHD simulation of a self-gravitating cloud performed with the AMR RAMSES code. The initial conditions of the simulation consisted of a 10\,000 M$_{\odot}$ ellipsoidal cloud with an aspect ratio of 2 and a density profile as $\rho(r,z)=\rho_0/[1+(r/r_0)^2+(z/z_0)^2]$, where $r=\sqrt{(x^2+y^2)}$, $z_0=2r_0$, $r_0=5$~pc, and $\rho_0=500$~cm$^{-3}$. The density at the edge of the cloud is $\rho_{edge}=50$~cm$^{-3}$, and is maintained in pressure equilibrium with an external medium at lower density. The magnetic field is perpendicular to the main axis of the cloud, with an intensity proportional to the cloud column density and a peak value of 7~$\mu$G. By the time of the snapshot presented in this paper, the magnetic field  had increased to $\sim100~\mu$G in the densest regions, with an average value over the dense gas of $\sim 20~\mu$G. The simulation is isothermal at a temperature of 10~K. A turbulent velocity field was seeded to initially obtain  $2T+W+M \simeq 0$ over the entire cloud, where $T$ is the kinetic energy (thermal and turbulent), $W$ the gravitational energy, and $M$ the magnetic energy. Turbulence was undriven and allowed to decay. These conditions lead to $W\simeq 2 T \simeq 9M$. Although globally, the turbulent and magnetic energy compensate for the gravitational energy of the cloud, it quickly becomes sub-virial due to the compressive nature of turbulence and because its energy quickly dissipates. The consequence of this is the fragmentation and global collapse of the simulated cloud.

\end{document}